\documentclass[sigconf]{acmart}

\AtBeginDocument{%
  \providecommand\BibTeX{{%
    \normalfont B\kern-0.5em{\scshape i\kern-0.25em b}\kern-0.8em\TeX}}}

\copyrightyear{2024}
\acmYear{2024}
\setcopyright{acmlicensed}\acmConference[CHI '24]{Proceedings of the CHI Conference on Human Factors in Computing Systems}{May 11--16, 2024}{Honolulu, HI, USA}
\acmBooktitle{Proceedings of the CHI Conference on Human Factors in Computing Systems (CHI '24), May 11--16, 2024, Honolulu, HI, USA}
\acmDOI{10.1145/3613904.3642044}
\acmISBN{979-8-4007-0330-0/24/05}

\usepackage{multirow}
\usepackage{bm}

\usepackage{listings}
\usepackage{tabularray}
\usepackage{CJK}

\usepackage{subcaption}
\usepackage{tabularx}
\usepackage{xcolor}
\usepackage{soul}
\usepackage{graphicx}
\usepackage{float}

\newcommand{\eg}{{e.g.,\ }}

\newcommand{\ie}{{i.e.,\ }}

\newcommand{\aka}{{a.k.a.,\ }}

\definecolor{oxfordblue}{rgb}{0.0, 0.13, 0.28}
\definecolor{harvardcrimson}{rgb}{0.79, 0.0, 0.09}
\definecolor{dartmouthgreen}{rgb}{0.05, 0.5, 0.06}
\definecolor{princetonorange}{rgb}{1.0, 0.56, 0.0}
\definecolor{yaleblue}{rgb}{0.06, 0.3, 0.57}
\definecolor{usccardinal}{rgb}{0.6, 0.0, 0.0}
\definecolor{uclablue}{rgb}{0.33, 0.41, 0.58}
\definecolor{msugreen}{rgb}{0.09, 0.27, 0.23}
\definecolor{cornellred}{rgb}{0.7, 0.11, 0.11}
\definecolor{pomegranate}{RGB}{192, 57, 43}
\definecolor{anti-pomegranate}{RGB}{43,178,192}
\definecolor{alizarin}{RGB}{231, 76, 60}
\definecolor{anti-belize}{RGB}{185, 41, 56}
\definecolor{belize}{RGB}{41, 128, 185}
\definecolor{peter}{RGB}{52, 152, 219}
\definecolor{green}{RGB}{22, 160, 133}
\definecolor{anti-green}{RGB}{160,22,118}
\definecolor{turquoise}{RGB}{26, 188, 156}
\definecolor{pumpkin}{RGB}{211, 84, 0}
\definecolor{anti-pumpkin}{RGB}{0,22,211}
\definecolor{carrot}{RGB}{230, 126, 34}
\definecolor{wisteria}{RGB}{142, 68, 173}
\definecolor{anti-wisteria}{RGB}{99,173,68}
\definecolor{amethyst}{RGB}{155, 89, 182}
\definecolor{nephritis}{RGB}{39, 174, 96}
\definecolor{anti-nephritis}{RGB}{174,39,117}
\definecolor{grey-bg}{RGB}{242,242,242}

\newcommand{\correct}[1]{{#1}}

\begin{document}
\begin{CJK*}{UTF8}{gbsn}

\title[Unpacking Chinese Adolescent Smartwatch-Mediated Socialization]{Wrist-bound Guanxi, Jiazu, and Kuolie: Unpacking Chinese Adolescent Smartwatch-Mediated Socialization}

\author{Lanjing Liu}
\authornote{The first two authors contributed equally to this work.}
\email{lanjing@vt.edu}
\orcid{0000-0003-2723-722X}
\affiliation{%
 \institution{Virginia Tech}
 \city{Blacksburg, Virginia}
 \country{USA}}

\author{Chao Zhang}
\authornotemark[1]
\email{cz468@cornell.edu}
\orcid{0000-0003-4286-8468}
\affiliation{%
 \institution{Cornell University}
 \city{Ithaca, NY}
 \country{USA}}

\author{Zhicong Lu}
\email{zhicong.lu@cityu.edu.hk}
\orcid{0000-0002-7761-6351}
\affiliation{%
 \institution{City University of Hong Kong}
 \city{Hong Kong SAR}
 \country{China}}

\begin{abstract}

Adolescent peer relationships, essential for their development, are increasingly mediated by digital technologies. 
As this trend continues, wearable devices, especially smartwatches tailored for adolescents, is reshaping their socialization.
In China, smartwatches like XTC have gained wide popularity, introducing unique features such as ``\textit{Bump-to-Connect}'' and exclusive social platforms.
Nonetheless, how these devices influence adolescents' peer experience remains unknown.
Addressing this, we interviewed 18 Chinese adolescents (age: 11---16), discovering a smartwatch-mediated social ecosystem.
Our findings highlight the ice-breaking role of smartwatches in friendship initiation and their use for secret messaging with local peers. 
Within the online smartwatch community, peer status is determined by likes and visibility, leading to diverse pursuit activities (\ie \textit{chu guanxi}, \textit{jiazu}, \textit{kuolie}) and negative social dynamics.
We discuss the core affordances of smartwatches and Chinese cultural factors that influence adolescent social behavior, and offer implications for designing future wearables that responsibly and safely support adolescent socialization.

\end{abstract}



\maketitle
\section{Introduction}

``\textit{Are you my buddy?}'' seems to be one of the most important questions among adolescents, underlining the significance of forming and nurturing peer relationships during these formative years~\cite{hayPeerRelationsChildhood2004a,rubinPeerRelationshipsChildhood2011}. 
In the modern era, the ubiquitous presence of personal devices, coupled with the ease of online connectivity, has transformed the ways in which adolescents initiate and maintain friendships~\cite{ledbetterCommunicationTechnologyInterpersonal2017,punamakiAssociationsInformationCommunication2009,shyamInformationTechnologyInternet2011}.
Over the years, Human-Computer Interaction~(HCI) researchers have revealed that devices such as smartphones, in conjunction with platforms like Facebook and Instagram, have been integrated into the fabric of adolescent social experiences, providing fresh pathways for connections~\cite{nesi_transformation_2018-1,nesi_transformation_2018-2,sunExploratoryStudyYoung2020,yaroshYouthTubeYouthVideo2016}.\looseness=-1


In the digital era, smartwatches, especially those tailored for the younger demographic, are gaining traction. 
The market value for child-oriented wearables is expected to rise from \$0.8 billion in 2023 to \$2.2 billion by 2030~\cite{GlobalKidsSmartwatchb,kidssmartwatchmarket}. 
Notably, smartwatches like ``小天才~(\textit{XTC\footnote{XTC stands for Xiaotiancai, the pinyin for ``小天才''.}})~\cite{XiaoTianCaiXiaoTianCaiGuanFangWangZhan}'' are popular among Chinese adolescents due to tailored social features. 
Despite a screen under two inches, these watches have a dedicated social network, accessible only to adolescent users of the same brand, ensuring a closely-knit community by excluding potential threats and even parents.

Being worn on the wrist, this smartwatch offers adolescents instant and continuous access to their digital world and allows them to stay connected with their peers ``\textit{anytime and anywhere}~\cite{barfieldBasicConceptsWearable2001}'', unlike smartphones that might be pocketed or taken away by caregivers~\cite{pizzaSmartwatchVivo2016,cecchinatoAlwaysLineUser2017}.
Furthermore, the inherent values of collectivism and interconnectedness rooted in the Eastern cultures~\cite{chuSocialCapitalSelfpresentation2010,yangVirtualGiftsGuanxi2011} may nuance how Chinese adolescents engage with and utilize their smartwatches for social purposes.
However, most existing research is either centered on adult smartwatch use~\cite{hanselPotentialWearableTechnology2018,liuAnimoSharingBiosignals2019}, or focuses on the online social behaviors of Western adolescents~\cite{nesi_transformation_2018-1,nesi_transformation_2018-2}.
This leaves a significant gap in understanding the role of smartwatches in adolescent socializing, the impact on their peer relationships, and the cultural dimensions that nuance these interactions. 
Investigating this area can enrich our comprehension of adolescents' social needs concerning smartwatches and complement the Western-centered perspective in earlier works by integrating a Chinese cultural lens. 
Such knowledge can also be instrumental in shaping the design of next-generation wearables and their social functionalities to foster adolescent social development.  
Hence, we are motivated to explore the subsequent research questions:\looseness=-1

\begin{itemize}
    \item[\textit{\textbf{RQ1:}}] \textit{How do Chinese adolescents use smartwatches to communicate and socialize with their peers?}
    \item[\textit{\textbf{RQ2:}}] \textit{How do cultural factors influence the ways Chinese adolescents engage in smartwatch-mediated socialization?}
    \item[\textit{\textbf{RQ3:}}] \textit{What are the perceived benefits and challenges of smartwatch-mediated socialization for Chinese adolescents?}
\end{itemize}

To address these research questions, we conducted in-depth semi-structured interviews with 18 Chinese adolescents~(aged 11-16 years) who have diverse experiences using smartwatches to socialize with their peers. 
We uncovered a comprehensive social ecosystem mediated by smartwatches, which encompasses initiating friendships and communicating with local peers, understanding group dynamics and social interactions with online peers, recognizing emerging negative social dynamics, and transitioning in and out of the smartwatch platforms.
We also found that the constant presence of smartwatches on the wrist serves as an icebreaker when initiating friendships with local peers. 
After making friends, adolescents use smartwatches for secret messaging, such as sharing homework answers.
In the online community within smartwatches, peer status is influenced by the number of likes and the extent of visibility. 
Chinese adolescents seek such status by building virtual relationships, establishing or joining peer groups, and expanding their circle of active friends.
Concurrently, negative social dynamics such as discrimination, drama, cyberbullying, and flame war have emerged.
Drawing on our findings, we discuss three core affordances~(i.e., physicality, locality, exclusivity) of smartwatches and Chinese cultural factors that influence adolescent social behavior.
\looseness=-1

In conclusion, this work contributes to HCI and CSCW by:~(i) uncovering a holistic social ecosystem mediated by smartwatches, elucidating the dynamics of forming friendships, understanding group dynamics, and recognizing both positive and negative social interactions, thereby offering insights into the nuanced role of wearables in adolescent socialization;~(ii) delving into the nuances of how Eastern cultural values of collectivism and interconnectedness shape and influence the ways Chinese adolescents engage with and utilize their smartwatches for socialization; and~(iii) offering design implications, aiming to guide the creation of future wearables that responsibly and safely support adolescent peer socialization.\looseness=-1
\section{Related Work}
We situated our study within previous research related to~(1) peer relations in adolescence,~(2) adolescent online social behavior,~(3) smartwatch use among children and adolescents, and~(4) online social interactions in Chinese contexts.


\subsection{Peer Relations in Adolescence}
Adolescence represents an essential developmental phase marked by a pronounced dependence on peer socialization, exerting a profound influence on personal development and psychological welfare~\cite{almquist_peer_2009, almquist_social_2013, menting_cognitive_2016, modin_childhood_2011}. 
Past research has investigated peer relations among adolescents~\cite{choukas-bradley_peer_2014, furman_friendships_2015, prinstein_peer_2016, rubin_children_2015}, with a focus on three domains: peer victimization, peer status, and peer influence~\cite{choukas-bradley_peer_2014, furman_friendships_2015, prinstein_peer_2016}.

Adolescents' socialization is shaped by their peer status, including peer acceptance, peer rejection, and peer popularity~\cite{choukas-bradley_peer_2014, prinstein_peer_2016, nesi_transformation_2018-2}. 
Peer popularity, denoting prominence and dominance within the peer hierarchy, represents a distinctive manifestation of social status, signifying leadership, influence, and esteem~\cite{cillessen_understanding_2005}.
During adolescence, the significance of peer popularity becomes particularly pronounced, which aligns with the heightened receptivity of adolescents to peer feedback and their considerations of social status~\cite{salmivalli_bullying_2010}. 
To attain peer status, adolescents employ various strategies, including bullying behaviors~\cite{caravita_agentic_2012, volk_adolescent_2015}, social comparison, and soliciting feedback from peers~\cite{borelli_reciprocal_2006, butzer_relationships_2006}. 
The pursuit of peer status transitions to the digital landscape. 
Online platforms magnify the essence of peer status through quantifiable metrics such as followers, comments, and likes~\cite{nesi_transformation_2018-2}, provide an avenue for any adolescent to achieve a semblance of ``celebrity'' status~\cite{marwick_its_2014}.

Peer victimization involves both attacking others and being attacked~\cite{juvonen_peer_2001} and occurs in widespread contexts, such as in and out of school, online platforms~\cite{troop-gordon_peer_2017}. 
Some adolescents derive enjoyment from peer victimization~\cite{smith_cyberbullying_2008}, some gain membership within a clique ~\cite{farmer_aggression_2007, troop-gordon_peer_2017}. 
Online peer victimization, also known as ``cyberbullying'', is an extension of traditional forms, often involving verbal aggression, social exclusion, and rumor dissemination~\cite{wang_patterns_2012}. 
Online platforms, with features like availability, visuality, publicness, permanence, and quantifiability, amplify cyberbullying's impact on adolescents compared to offline victimization. 
For example, availability enables harm from anywhere, reaching large audiences easily, and publicness and anonymity affect the perceived severity of bullying~\cite{sticca_is_2013}.


Peer influence is notably significant at the adolescent stage~\cite{brechwald_beyond_2011}, because of increased identity exploration and expanded unsupervised peer interactions~\cite{prinstein_peer_2016}. 
Peer influence covers risk-taking behavior~\cite{somerville_dissecting_2019, loke_family_2013}, prosocial behavior~\cite{choukas-bradley_peer_2015}, decision-making~\cite{choukas-bradley_peer_2014}, social acceptance~\cite{veenstra_prominence_2021}, and academic performance~\cite{temitope_influence_2015}. 
Online platforms have magnified peer influence from speed, volume, and scope~\cite{nesi_transformation_2018-2} by facilitating adolescents' access to a wider range of peers' personalities, opinions, and behaviors~\cite{courtois_social_2012}.

Adolescents' peer victimization, peer status, and peer influence exhibit intricate interrelations~\cite{dijkstra_beyond_2008, cillessen_stability_2015, cook_predictors_2010, troop-gordon_peer_2017}.
Past research has shown that peer victimization, peer status, and peer influence among adolescents persist in computer-based and smartwatch-based socialization. 
Digital platforms have strengthened these relationships in terms of frequency, duration, scope, and influence. 
In comparison to computers and smartphones, smartwatches not only enrich adolescents' online interactions but also intertwine with their offline social engagements. 
However, there exists a research need in exploring the influence of smartwatches on these peer relationships.
Therefore, this paper focuses on the influence of smartwatches on peer relationships among adolescents, and compares and contrasts this with computer-based interactions.

\subsection{Adolescent Online Social Behavior}

Online environment has influenced adolescents' socialization~\cite{nesi_transformation_2018-2}, extended their communication with friends~\cite{valkenburg_online_2011} by nearly constant opportunities for interaction~\cite{spies_shapiro_growing_2014}.
Adolescences transpose conventional offline social norms and behaviors into their online interactions~\cite{subrahmanyam_connecting_2006}, grape with issues and challenges across both spheres~\cite{olweus_cyberbullying_2012}, and develop broader social competence compared to offline interactions~\cite{reich_connecting_2017}.
Additionally, adolescents tend to shape their offline actions based on their online performances, becoming increasingly conscious of how their offline experiences of friendships will be projected to peers through social media posts, pictures, and comments~\cite{nesi_transformation_2018-2}.

Throughout socialization online, adolescents initiate their socialization from a personal standpoint, such as identity exploration~\cite{ok_be_2021}, self-presentation~\cite{mcroberts_it_2016,yarosh_youthtube_2016}, and emotional disclosure. 
The potential for quantifiable, immediate responses from peers makes teens more actively engaged in these social activities online~\cite{nesi_transformation_2018-1}.
Then adolescents nurture existing connections~\cite{reich_friending_2012,scott_contemporary_2021}, make new friends~\cite{fields_i_2015}, establish social group or cliques~\cite{li_i_2023,desjarlais_socially_2017}.
Online platforms empower adolescents in their socialization by granting control and autonomy~\cite{valkenburg_preadolescents_2007c, young_cognitive_2012}, access to like-minded peers~\cite{subrahmanyam_digital_2011,love_exploring_2012,stewart_brief_2011}.
Also, the digital landscape introduces new social behavior such as ``drama,'' characterized by performative conflicts on social media platforms~\cite{marwick_its_2014}. 
Consequently, adolescents' perception is one of a broader and more profound range of communication topics when interacting online~\cite{peter_research_2006,schouten_precursors_2007,valkenburg_internet_2007a,valkenburg_preadolescents_2007c}.
Adolescents' social behaviors also contain some negative ones, including bullying~\cite{ojanen_instrumental_2013,troop-gordon_peer_2017}, harassment~\cite{ali_understanding_2022}, and relational aggression~\cite{wang_patterns_2012}.
Additionally, the digital environment makes aggressive and harassing messages easy for adolescents, which is often underreported due to adolescents' lower readiness for defense compared to adults~\cite{agatston_students_2007,smith_cyberbullying_2008,draucker_role_2010}.

HCI community has explored adolescent online social interactions extensively~\cite{ashktorab_designing_2016}. 
Building on this, efforts have been made to enhance online social environments for adolescents. 
For instance, providing a safe online socializing environment for kids and teens through physical restrictions~\cite{cheok_petimo_2009}, fostering positive online behavior and mitigating cyberbullying~\cite{fan_feelbook_2016}. 
These endeavors also aim to aid adolescents in reflecting upon their use of social media platforms~\cite{davis_supporting_2023}.
Online platforms are crucial for adolescent socialization, providing increased social opportunities and experiences. While past studies primarily examined social behavior on traditional platforms through computers or phones, today's adolescents are increasingly using smartwatches for social interactions.
However, the specifics of adolescent social interactions on smartwatches remain largely unexplored. 
Compared to computers or phones, smartwatches feature smaller screens and offer more discreet and limited social interactions.
This paper investigates how adolescents engage in social interactions using smartwatches and proposes design implementations.

\subsection{Smartwatch Use among Children and Adolescents}

Smartwatches, as described by Barfield and Caudell~\cite{barfieldBasicConceptsWearable2001}, are ``\textit{fully functional, self-powered, self-contained computers worn on the body, offering access to, and interaction with, information anytime and anywhere}''.
Their persistent presence on the wrist makes them ``context-insensitive,'' requiring only a familiar gesture for activation~\cite{pizzaSmartwatchVivo2016}.
Given concerns over smartphones, many parents view smartwatches as a preferable alternative~\cite{Wait8th}.
These wearables offer essential communication and locational features, allowing parents to maintain contact with their children and ensuring their safety.
In China, this preference is even more evident, partly because of the strict ``minor model\footnote{The ``minor mode'' regulations proposed by the Chinese government prevent internet addiction by limiting children younger than 8 to 40 minutes of smartphone time a day. The time limit would increase with age, reaching two hours daily for those ages 16 to 18.}'' policy.
Currently, about one-third of Chinese children own a smartwatch~\cite{chinareporthall20222027NianZhongGuoErTongZhiNengShouBiaoXingYeShenDuDiaoYanJiTouZiQianJingYuCeYanJiuBaoGao2022, chinaindustrialresearchnetwork20222027NianZhongGuoErTongZhiNengShouBiaoXingYeFaZhanQuShiJiTouZiFengXianYanJiuBaoGao2022}.
The most popular smartwatch is XTC~\cite{XiaoTianCaiXiaoTianCaiGuanFangWangZhan}.
It offers distinctive social features beyond the essentials, like instant massaging and social networks~(discussed in Section~\ref{sec:research_context}).

Studies on smartwatch usage among children began with Williams et al.'s work~\cite{williamsWearableComputingGeographies2003} in 2003, which involved co-designing soundscape wearable applications with children. 
Then, numerous scholars have examined children's adoption of smartwatches for fitness monitoring~\cite{ankrahMeMyHealth2022}, activity tracking~\cite{oygurLivedExperienceChildOwned2021}, and self-regulation~\cite{cibrianBalancingCaregiversChildren2019, silvaUnpackingLivedExperiences2023}, mostly focusing on health informatics~\cite{amiriWearSenseDetectingAutism2017, capodieciDynamicallyAdaptingEnvironment2018, jeongSmartwatchWearingBehavior2017a, pizzaSmartwatchVivo2016, shihUseAdoptionChallenges2015, vaizmanRecognizingDetailedHuman2017}.
Additionally, there is a distinct strand of research centered on the implementation of smartwatches in educational contexts~\cite{bowerWhatAreEducational2015a, dearriba-perezUseCommercialWrist2017, parkDesignWearableSensor2002, schollWearablesWetLab2015}, such as scientific learning~\cite{chuUnderstandingContextChildren2019, chuWearableAppDesign2017, garciaWearablesLearningExamining2018} and fostering creativity~\cite{vishkaieCanWearableTechnology2018, kazemitabaarMakerWearTangibleApproach2017}.

Despite the increasing popularity of smartwatches among children and the introduction of built-in social features, there is a significant research gap in empirically understanding children's social behavior with respect to these devices.
While the always-accessible nature of smartwatches offers potential benefits, such as enabling children to maintain connections with peers during transitions~(e.g., moving from home to school or outdoors)~\cite{cecchinatoAlwaysLineUser2017, pizzaSmartwatchVivo2016}, only a handful of studies have delved into adults' use of smartwatches for lightweight social connection~\cite{liuAnimoSharingBiosignals2019} and social interaction monitoring~\cite{hanselPotentialWearableTechnology2018}.
Even though these studies were not specifically tailored for children, their outcomes underscore the potential of smartwatches in mediating children's social interactions and experiences with peers. 
For example, much of communication online via smartwatches contains non-textual cues like emojis and biosignals~\cite{hanselPotentialWearableTechnology2018}.
Even ``one-click communication,'' such as ``liking'' online social media posts, can provide diverse cues~(\eg support, agreement)~\cite{scissorsWhatAttitudesBehaviors2016}.
To fill this gap, this paper investigates how Chinese adolescents use XTC, the most popular children's smartwatch in China, to engage in social interactions with their peers and understand the nuances of their social behaviors.

\subsection{Online Social Interactions in Chinese Contexts}

The behavior of adolescents, especially in the realm of social interactions, is significantly shaped by the cultural milieu they inhabit~\cite{brizioNoMoreChild2015, liebkindAcculturationExperienceAttitudes2022}. 
Delving deep into the inter-cultural dynamics reveals profound differences in social orientations between the West and East. 
Central to inter-cultural sociology and cultural psychology, Western societies predominantly exhibit values rooted in independence, individualism, autonomy, and self-achievement~\cite{hofstedeCultureConsequencesInternational1984}, seeing the self as distinct and separate from others.
Conversely, East Asian cultures prioritize principles of interdependence, harmony, relatedness, and connection~\cite{hofstedeDimensionsNationalCultures1983, singelisMeasurementIndependentInterdependent1994, triandisIndividualismCollectivism2018}. 
These societies, being inherently interdependent-oriented, view the self not as an isolated entity but as an integral component of a broader network of meaningful social relationships~\cite{markusCultureSelfImplications2014}. 
This foundational understanding of the self in Asian contexts is emblematic of their emphasis on belonging. 
Here, individuals are often tasked with fulfilling a myriad of obligations and responsibilities towards others, which form the bedrock of their societal interactions~\cite{heineThereUniversalNeed1999}. 

Focusing specifically on the Chinese scenario, the profound influence of the collectivistic culture, deeply ingrained in Confucianism, becomes evident. 
Chinese social media users harness their memberships and relationships as facets to forge their identities~\cite{wangWeChatMomentsInternational2023}.
This not only allows them to anchor their position within online spheres but also imparts an enhanced sense of belonging~\cite{chuSocialCapitalSelfpresentation2010}.
Another element that colors the fabric of social interactions in China is the practice of ``关系~(\textit{guanxi})''. 
\textit{Guanxi}, a uniquely Chinese form of social exchange, plays a crucial role in online communities~\cite{yangVirtualGiftsGuanxi2011}.
At its core, \textit{guanxi} pertains to the relational ties established between an individual and others~\cite{jacobsConceptGuanxiLocal1982}, often nurtured through a continuous exchange of favors with the intent to foster harmonious relations~\cite{pyeChineseCommercialNegotiating1982}.
Interestingly, while the nuances of \textit{guanxi} in general Chinese online communities are well-documented, its specific manifestations among Chinese adolescents remain unexplored. 
Given the pervasiveness of this cultural norm, it is reasonable to infer that their online social behaviors may also be influenced by \textit{guanxi}. 

Addressing the Western-centric bias, a growing body of HCI research has ventured into understanding the specific cultural nuances that characterize technology-mediated communication in China~\cite{luYouWatchYou2018, heSeekingLoveCompanionship2023, luFeelItMy2019, tangDareDreamDare2022,luFifteenSecondsFame2019, chenMyCultureMy,caoYouRecommendBuy2021}.
However, the current body of research on children's technology-mediated socialization is heavily skewed towards Western contexts. 
While such a focus has undeniably enriched the literature with invaluable insights, it may overlook the multifaceted social behaviors exhibited by children in non-Western settings like China.
Our paper seeks to address this by revealing the intricate interplay between cultural norms and smartwatch-mediated socialization among Chinese adolescents.





\section{Research Context: Smartwatches for Chinese Youth}
\label{sec:research_context}

\begin{figure*}
 \includegraphics[width=\linewidth]{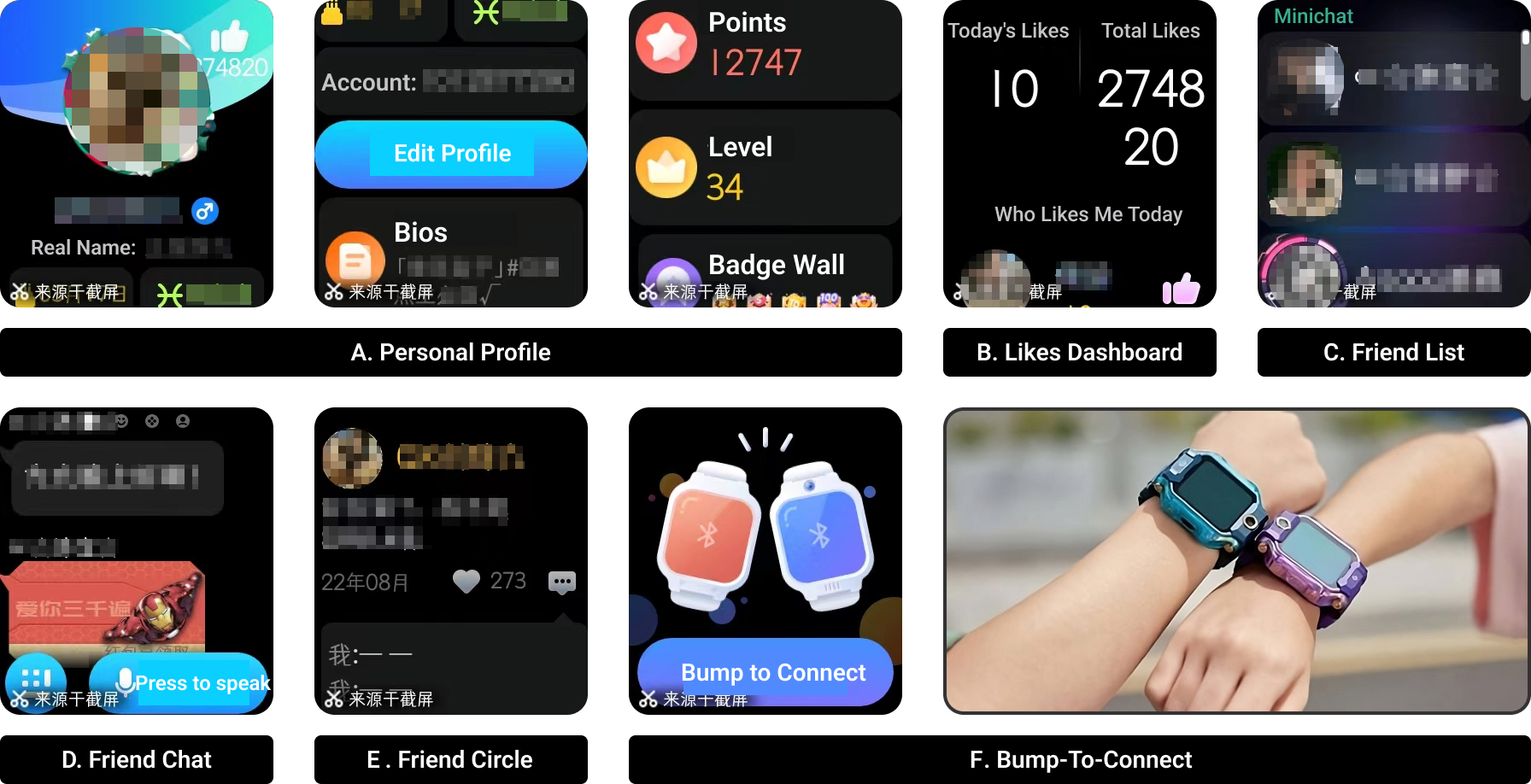}
 \caption{The main interface, features, and appearance of smartwatches.
 A. Personal Profile, including profile picture, nickname, real name, account information, bios, points, level, and badge wall.
 B. Dashboard for likes, featuring 'today's likes,' 'total likes,' and 'who likes me today.'
 C. Friend List within MiniChat, displaying friends' profile pictures and nicknames.
 D. Friend Chat functionality allowing users to input text via voice and send 'points' red packets, among other features.
 E. Friend Circle, a platform for users to post content and receive likes and comments.
 F. Bump-To-Connect feature interface and user gestures.}
 \Description{The main interface, features, and appearance of smartwatches.
 A. Personal Profile, including profile picture, nickname, real name, account information, bios, points, level, and badge wall.
 B. Dashboard for likes, featuring 'today's likes,' 'total likes,' and 'who likes me today.'
 C. Friend List within MiniChat, displaying friends' profile pictures and nicknames.
 D. Friend Chat functionality allowing users to input text via voice and send 'points' red packets, among other features.
 E. Friend Circle, a platform for users to post content and receive likes and comments.
 F. Bump-To-Connect feature interface and user gestures.}
 \label{fig:user_interface}
\vspace*{-6pt}
\end{figure*}

In this paper, we chose XTC as our research context, aligning with a prevalent approach seen in many HCI interview-based papers that focus on popular platforms~\cite{DMS,bilibili,shen_labor_2021}. 
This research approach deepens our understanding of emerging and unexplored phenomena or domains.

The children's smartwatch industry in China began in the mid-2010s, initially focusing on tracking a child's location for safety. 
In 2015, ``小天才~(\textit{XTC})'', launched their first generation of smartwatches. 
Beyond location tracking, it offered phone calling, instant messaging, and the innovative ``碰一碰~(\textit{Bump-to-Connect})'' feature~(Figure~\ref{fig:user_interface}F). 
It allowed children to touch their watches together, adding each other to their contact lists, transforming the watch from a mere safety device to a social tool. 
This shift contributed to its surge in popularity, resulting in about 30\% market penetration among Chinese youth~\cite{chinareporthall20222027NianZhongGuoErTongZhiNengShouBiaoXingYeShenDuDiaoYanJiTouZiQianJingYuCeYanJiuBaoGao2022,chinaindustrialresearchnetwork20222027NianZhongGuoErTongZhiNengShouBiaoXingYeFaZhanQuShiJiTouZiFengXianYanJiuBaoGao2022}.

As of now, XTC stands as the most popular smartwatch brand in China, 
as the first with 35\% share and more than twice as much as the second place~\cite{limSmartwatchShipmentForecast2023}.
XTC tailored for adolescents with various social features.
One of its notable functions is the social networking platform ``微聊~(\textit{MiniChat})''(Figure~\ref{fig:user_interface}C). 
Within \textit{MiniChat}, each child is provided with a unique profile displaying their avatar, username, real name, gender, birthday, astrological sign, and bio, all of which can be customized~(Figure~\ref{fig:user_interface}A).
Profiles also exhibit a ``Like'' counter, user levels, and badges. 
Users' ability to give Likes to another profile is tiered based on their levels\footnote{Those between levels 1--10 can offer up to 5 Likes, levels 11--20 can give up to 10, and only those above level 20 can give a maximum of 20 Likes.}.
User levels advance through daily rewards and purchases using points. 
These points, which act as the platform's virtual currency, are earned through active participation and can be used to buy virtual items or even be gifted between friends.
In addition, users can also view the number of likes received today, the total number of likes received, and the list of users who liked today in the dashboard~(Figure~\ref{fig:user_interface}B).\looseness=-1

Regarding connectivity, \textit{MiniChat} allows users to add up to 150 friends via methods like \textit{Bump-to-Connect} or phone number searches. 
Once friends, children can send messages~(Figure~\ref{fig:user_interface}D), share media~(Figure~\ref{fig:user_interface}E), or make calls. 
The platform also permits users to create 2 or join 5 group chats, with a capacity of 11 members each. 
Group chats further streamline friend additions by enabling profile checks of other participants.
Additionally, \textit{MiniChat} boasts a feature called ``好友圈~(\textit{Friend Circle})'', akin to social media platforms like Facebook or Instagram, where users can post updates, photos, and videos for their friends to view, like, and comment on~(Figure~\ref{fig:user_interface}E).
It is worth noting that both \textit{MiniChat} and \textit{Friend Circle} are exclusive, permitting access only to fellow users of the XTC smartwatches, creating a closely-knit community of young users, while intentionally excluding potential threats such as predators, as well as parents.


Furthermore, XTC offers a mobile app for parents, allowing them to communicate with their child, monitor social activities, set usage limits, and track the child's location and health data. 
Importantly, parents cannot access the content of their child's conversations or posts in the \textit{Friend Circle} on the watch. 
While this parent app is part of the XTC ecosystem, our paper primarily focuses on children's use of smartwatches for socialization.\looseness=-1
\section{Methods}
To comprehend smartwatch-mediated socialization among Chinese adolescents and their corresponding experiences and perceptions, we conducted a semi-structured interview study.
Our interview involved 18 participants from June to August 2023, surpassing 12, the most common sample size in interview-based studies of CHI~\cite{caine_local_2016}. 
In HCI community, self-reporting interview has also been successfully, widely applied in many prior studies of child-computer interaction~\cite{freed_understanding_2023,sun_child_2021,xiao_sensemaking_2022,pitt_kids_2021}. 
Compared to quantitative research, qualitative research is flexible in emerging issues and emphasizes a deep and ``thick'' understanding of the diversity and complexity of personal experiences and feelings, and social behaviors~\cite{Qualitative_Data}.\looseness=-1

\subsection{Participants}

\begin{table*}
    \caption{Basic information summary of participants interviewed. Location: Cities ranked as per the Chinese city tier system~\cite{wikipediaChineseCityTier2023}~(All cities in China are categorized from Tier 1~(highest) to Tier 4~(lowest), following the same classification system of prior HCI studies in Chinese contexts~\cite{heSeekingLoveCompanionship2023}. It is noted that we reported the city tier based on the participants' location of the city; in fact, five participants were from towns or rural areas affiliated with the city). \# Contacts: Number of contacts the participant has on their smartwatch at the time of interview. The upper limit is 150. \# Likes: Number of profile likes the participant had received up to the interview date. Have Phone?: Indicates if the participant owns a phone. If affirmative, it further specifies if they always have access to it or only during vacations. \textit{Biaoquan}?: Denotes if the participant is involved in \textit{biaoquan}, actively seeking profile likes and visibility. If yes, it notes whether they are a \textit{dalao}, characterized by having over 200k profile likes and a notable visibility in \textit{biaoquan}. These concepts are displayed in Table~\ref{tab:concepts} and elaborated in Section~\ref{sec:online_peers}.
    Male participants reported an average weekly usage of 6.28 hours, whereas female participants reported a higher usage of 16.5 hours. These findings align with prior research results~\cite{shah_high_2023,b_social_2023,mohd_reffal_users_2022}.}
    
    \label{tab:demographics}
    \begin{tabular}{lllllllllll}
    \toprule
    \textbf{ID} & \textbf{Gender} & \textbf{Age} & \textbf{Location} & \textbf{Wkly Use} & \textbf{Yrs Use} & \textbf{\# Contacts} & \textbf{\# Likes} & \textbf{Have Phones?} & \textbf{\textit{Biaoquan}?} \\
    \hline
    C01 & Male & 13 & Tier 4 & 2 hours & 5 years & 4 & $<$100 & Yes~(vacation) & No \\
    C02 & Female & 11 & Tier 4 & 10 hours & 1 year & 7 & $<$100 & Yes~(always) & No \\
    C03 & Female & 15 & Tier 4~(rural) & 15 hours & 7 years & 79 & 30k$+$ & Yes~(vacation) & Yes \\
    C04 & Male & 11 & Tier 2 & 7 hours & 1 year & 36 & $<$100 & No & No \\
    C05 & Male & 13 & Tier 4~(rural) & 12 hours & 6 years & 101 & 6k$+$ & Yes~(vacation) & No \\
    C06 & Male & 13 & Tier 3~(rural) & 10 hours & 6 years & 40 & 50k$+$ & Yes~(always) & Yes \\
    C07 & Female & 15 & Tier 1 & 28 hours & 5 years & 142 & 270k$+$ & Yes~(always) & Yes~(\textit{dalao}) \\
    C08 & Female & 13 & Tier 2 & 11 hours & 7 years & 91 & 10k$+$ & Yes~(vacation) & No \\
    C09 & Male & 14 & Tier 4 & 3 hours & 7 years & 122 & 8k$+$ & Yes~(vacation) & No \\
    C10 & Female & 16 & Tier 1 & 14 hours & 6 years & 23 & 120k$+$ & No & Yes \\
    C11 & Female & 14 & New Tier 1 & 7 hours & 5 years & 145 & 800k$+$ & Yes~(always) & Yes~(\textit{dalao}) \\
    C12 & Female & 14 & New Tier 1 & 30 hours & 4 years & 103 & 70k$+$ & Yes~(vacation) & Yes \\
    C13 & Male & 15 & New Tier 1 & 5 hours & 3 years & 44 & 280k$+$ & Yes~(vacation) & Yes~(\textit{dalao}) \\
    C14 & Female & 14 & Tier 2~(rural) & 7 hours & 1 year & 143 & 460k$+$ & Yes~(always) & Yes~(\textit{dalao}) \\
    C15 & Female & 13 & New Tier 1 & 14 hours & 5 years & 68 & 270k$+$ & Yes~(always) & Yes~(\textit{dalao}) \\
    C16 & Male & 11 & Tier 2~(rural) & 5 hours & 4 years & 130 & 100k$+$ & No & Yes \\
    C17 & Female & 12 & Tier 3 & 35 hours & 6 years & 84 & 60k$+$ & Yes~(always) & Yes \\
    C18 & Female & 12 & Tier 4 & 10 hours & 6 years & 112 & 40k$+$ & Yes~(always) & Yes \\
    \bottomrule
    \end{tabular}
\end{table*}

We defined our participant criteria as follows:
1) aged between 11 and 16 years; 
2) current ownership and use a smartwatch for over three months; and 
3) active engagement, meaning that they chat, post, collect and give likes via their smartwatch on a daily basis.
Our focus on ages 11 to 16 stems from:
1) adolescents at this age begin to learn and emulate adult-world interpersonal relationships and social order~\cite{kohlberg_psychology_1984}; and
2) the pivotal role of adolescent socialization in future achievements and well-being.
We employed various recruitment methods, including posting on social media platforms, direct messaging to the users, as well as snowball sampling. 
The participants' information is shown in Table~\ref{tab:demographics}.\looseness=-1

The 18 participants aged 11 to 16~(\(M = 13.28, SD = 1.45\)), consisting of 7 males and 11 females. 
They represented diverse urban contexts, encompassing tier 1 to tier 4 cities.
Their use of smartwatches and participation in smartwatch-based socialization varies.
Weekly smartwatch usage spanned 2 to 35 hours, with usage duration ranging from 1 to 7 years. 
Friend counts ranged from 4 to 145 and received likes from under 100 to 800,000. 
Notably, C07 suffers from the Asperger Syndrome. 
Although we reached theoretical saturation after 15 adolescent smartwatch users, we extended our study to include 18 participants to ensure solid foundation of results.\looseness=-1

\subsection{Data Collection}

\subsubsection{Semi-structured interviews}
Our semi-structured interviews were conducted in Mandarin, and encompassed the following sections:
The first section involves obtaining a fundamental understanding and overview of interviewee's smartwatch usage. 
This section primarily covers motivation to buy a smartwatch, frequently utilized features, and interviewee's overall perceptions of smartwatch usage, such as ``\textit{Why did you buy this smartwatch at the time?}'' and ``\textit{What do you normally use a smartwatch for?}''.
The second section proceeds to delve into adolescents' online socialization on smartwatches, exploring their engagement in online activities, experiences, and sentiments related to smartwatch-mediated social interactions. 
We ask questions like, ``\textit{Can you share an experience of making friends with someone on your smartwatch?}''and ``\textit{What do you think the likes are good for?}''.
The third section focuses on the impact of smartwatches on adolescents' offline social interactions, including potential effects on pre-existing friendships in the offline realm. We ask questions like, ``\textit{Do you think smartwatches have affected your real-life friendships?}''.
Lastly, we inquire about interviewee' comparative viewpoints regarding smartwatches and other smart devices~(\eg smartphones, tablets). 
Building upon this foundation, speculative design considerations are introduced, such as ``\textit{If you were the smartwatch designer, what kind of features would you like to design?}''.
The full list of interview questions is provided in ~\ref{appendix:interview_questions}.\looseness=-1 

The interviews were remotely conducted using video or audio calls. 
The interviews commenced by confirming interviewee' assent and their consent for recording. 
Next, we proceeded with the interview questions and followed up as needed. 
The interviews were conducted collaboratively by the first two authors: one interviewer primarily asked the questions while the other focused on note-taking. 
The interview duration ranged from 38 to 90 minutes with an average length of 74. 
In appreciation for their involvement, each participant was compensated with a 100 CNY~(approximately 15 USD) honorarium.\looseness=-1

\subsubsection{Smartwatch Content}
Furthermore, we augmented our insights by directly gathering supplementary data from participants. 
During interview, 12 participants showcased various smartwatch content, including its basic functions, personal profiles, and \textit{Friend Circle}.
9 participants provided more detailed content, including screenshots of features like personal profiles and chat records, as well as posters originating from smartwatch socialization and content directly related to smartwatch interactions on other platforms~(\eg Wechat\footnote{Wechat is a Chinese instant messaging, social media, and mobile payment app}, Xiaohongshu\footnote{Xiaohongshu is a Chinese social media and e-commerce platform. It sometimes referred to as ``Chinese Instagram''}, and Douyin\footnote{Douyin is a Chinese counterpart of TikTok}).
These data uniquely provide us with unfiltered glimpses into organic interactions, unaffected by researcher presence.\looseness=-1

\subsection{Data Analysis}
We collected comprehensive data for our analysis, consisting of 22-hour interview transcripts with detailed notes and smartwatch content from our participants.
We adopted an open coding method~\cite{corbin_basics_2015} to analyze both transcripts and the supplementary data. 
Two Chinese-speaking authors first coded 20\% of data independently and met to reach a consensus on the initial codes such as ``\textit{online activities within group chats}'' and ``\textit{criteria for adding peers to the friend list}''.
Two authors read the interview transcripts several times and coded a sample of the data collaboratively to code-book.
Using the initial code-book, authors coded the rest of the data individually. 
Two authors constantly compared and discussed their codes and resolved any disagreements as they coded, then updated the code-book as needed. 
Upon completion of the coding, two authors cross-checked each other's coding again to ensure full agreement. 
Then, all authors examined and discussed the code-book, and grouped the codes into higher-level themes.
The analysis was discussion-based, with no necessity for calculating inter-rater reliability~\cite{mcdonald2019reliability}.
After open coding, we used affinity diagramming~\cite{muller_curiosity_2014} to uncover main themes like ``\textit{Socializing with Online Peers}'' and sub-themes like ``\textit{Social Activity: Collect Like and Improve Visibility}''.
All researchers cross-referenced the original data, the code-book, and the themes, to make final adjustments. 
After multiple iterations, our analysis resulted in 4 themes and 20 sub-themes.
Rigorous discussions within the research team and reflective notes ensured the validation of our interpretations.


\subsection{Ethical Considerations}
Our study was approved by the last author's institute IRB. 
All parents provided informed consent while children gave their assent prior to the interview. 
Our method involved conducting semi-structured interviews with adolescents, which is similar to previous works~\cite{sun_they_2021,jin_exploring_2022}.
In addition, we took several measures to ensure the confidentiality of participants' sensitive personal information:
(1) before the interview, we extensively explained the research objectives, scope, participants' rights, and measures for ensuring privacy to both the participants and their guardians;
(2) during the interview, we refrained from directly or indirectly revealing the real identities of participants and any individuals they mentioned. Regarding sensitive issues, such as family matters, we avoided probing deeply to prevent emotional distress or discomfort;
(3) all data is securely stored, accessible exclusively to authorized researchers; 
(4) during the analysis and reporting of data, we merged or omitted details that might expose participants' identities, subjecting the materials provided by participants to de-identification processes; and
(5) all images presented in this paper have been redrawn, had sensitive information obscured, or had their resolution reduced to protect the privacy of our participants.
To address potential social desirability bias in self-report, we employed careful word and preface questions during interviews~\cite{latkin_relationship_2017}. 
Additionally, we engaged in selective self-disclosure to prompt a reciprocal exchange of information from interviewees, as suggested by prior research~\cite{orne_social_1962}.
These methods increased the depth and authenticity of the responses and fostered an open and frank conversation.

\subsection{Positionality Statement}
The authors conducting interviews and data analysis were born and raised in China. 
They bring a wealth of experience in interacting with Chinese adolescents at this age and possess a deep familiarity with various social patterns exhibited by Chinese adolescents. 
\section{Findings}
We described four major themes that captured how Chinese teens use smartwatches to socialize. 
We first describe smartwatch-based offline and online socialization activities in Section~\ref{sec:local_peers} and Section~\ref{sec:online_peers}.
In Section~\ref{sec:negative}, we unpacked negative social dynamics within smartwatches.
Finally, we showed the perceived benefits and challenges of smartwatch-based socialization in Section~\ref{sec:benefitsandchallenges}.

\begin{figure*}
 \includegraphics[width=\linewidth]{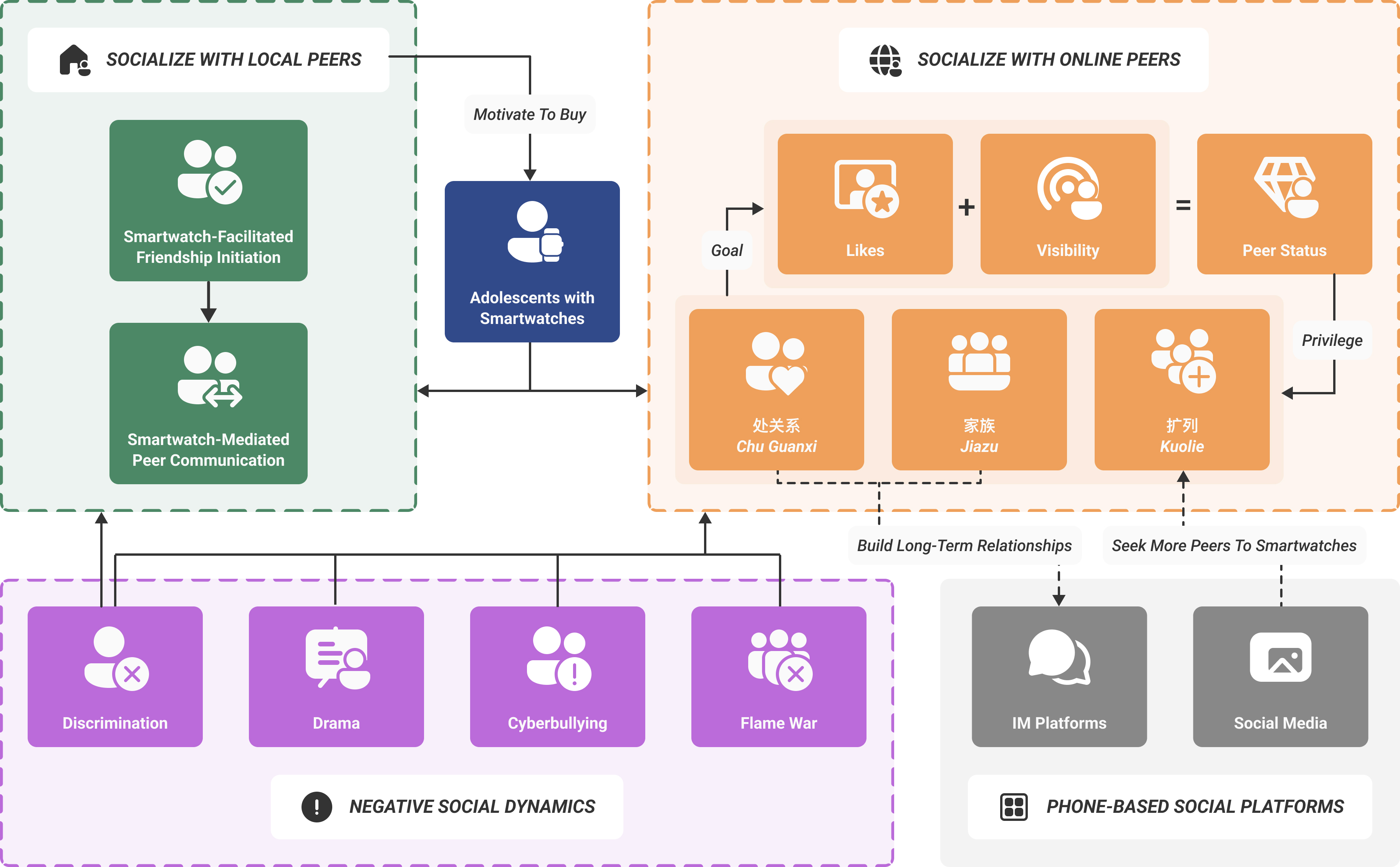}
 \caption{The social ecosystem of Smartwatch-Mediated Socialization, encompassing socialization with local peers, socialization with online peers, and negative social dynamics.}
 \Description{The social ecosystem of Smartwatch-Mediated Socialization, encompassing socialization with local peers, socialization with online peers, and negative social dynamics.}
 \label{fig:ecosystem}
\end{figure*}

\subsection{Socializing with Local Peers via Smartwatches}
\label{sec:local_peers}

The inherent presence of smartwatches on the wrist~\cite{pizzaSmartwatchVivo2016}, combined with the emerging \textit{Bump-to-Connect} interaction, underscores the potential of smartwatches to foster connections between local peers.
In our study, we found that the driving factor behind the widespread adoption of smartwatches among Chinese adolescents is rooted in their desire for enhanced offline social interactions.
C06 states, ``\textit{Almost everyone in my class has a XTC smartwatch. Some bought it just for making friends because many people use it.}'' 
C02 even persuaded their friends to buy the watch so they could chat without using their parents' phones. 
The device fosters ``\textit{better and faster relationships with friends}''~(\eg C04), offering convenience in staying connected~(\eg C05, C11) through social features and practicality~(\eg C12, C14). 

Furthermore, we discerned two predominant patterns of smartwatch-mediated socialization with local peers, as depicted in Figure~\ref{fig:ecosystem}: \textit{smartwatch-facilitated friendship initiation} and \textit{smartwatch-mediated peer communication}. 
Both of these cater to distinct social needs within real-life peer interactions.

\subsubsection{Smartwatch-Facilitated Friendship Initiation}

Our study reveals that the initiation of friendships is facilitated by the physical presence and the \textit{Bump-to-Connect} feature of XTC smartwatches. 
These devices not only act as conversation starters due to their tangibility but also provide an efficient and straightforward mechanism for adolescents to connect, turning everyday moments into opportunities for developing new friendships.

The physical presence of smartwatches on the wrist serves as ice-breakers in initiating interpersonal interactions among Chinese adolescents.
Participants often highlighted that the mere sight of a smartwatch on one's wrist often becomes a conversation starter. 
For instance, C04 mentioned how a simple observation about their watch led to a budding friendship, ``\textit{When they saw it, they said, `Hey, you have a watch?' Then the two of us connected.}''
Similarly, C01 and C06 emphasized the communal and serendipitous nature of such interactions by noting, ``\textit{Once I met a young guy on the basketball court in my neighborhood. He played basketball well. After the game, as he was leaving, he took out his watch, and I said `You also have a XTC watch!'. Then, we added each other as friends.}''~(C06).
Such experiences underscore the catalyst role of smartwatches in initiating interpersonal interactions and facilitating casual social interactions.

The \textit{Bump-to-Connect} feature embedded in these smartwatches appears to be a favorite among teenagers for establishing connections in offline scenarios. 
Many participants, like C01 and C04, expressed their preference for this feature, praising its straightforwardness and efficiency in connecting with others.
As C04 pointed out, ``\textit{I prefer Bump-to-Connect because it's relatively faster. It's quicker, and you can connect on the spot,}'' emphasizing the immediacy of the method. 
Additionally, the physicality of this interaction offers a layer of authenticity and trust, contrasting it with online interactions that might be more ambiguous, with C09 noting, ``\textit{When you Bump-to-Connect offline, you know what kind of person is on the other side, which is a bit safer.}'' 
As such, smartwatches, through their physical presence and features like \textit{Bump-to-Connect}, provide Chinese adolescents with versatile avenues for initiating friendships in varied offline contexts.\looseness=-1

\subsubsection{Smartwatch-Mediated Peer Communication}

After initiating friendships, smartwatches provide Chinese adolescents with a novel avenue for peer communication.
They communicate with their friends through voice input or typed input.
Same as the smartphone, the smartwatch serves as a bridge, fostering friendships and facilitating conversations that might not occur in face-to-face settings. 
C03 highlighted the case of a close friend who is typically reticent in person but becomes talkative on the smartwatch, illustrating how the device can empower quieter individuals to open up. 
C01 echoed this sentiment, emphasizing how smartwatches have ``\textit{strengthened friendships},'' allowing for more frequent and convenient communication, even over long distances. 
Interestingly, many of the participants emphasized the covert nature of smartwatch-mediated communication, with one participant noting how it allows for ``\textit{secret messaging}'' that escapes the notice of parents or teachers. 
For instance, C02 mentioned, ``\textit{In school, we avoid discussing certain topics like copying homework, fearing someone might overhear and report to the teacher. Instead, we use our smartwatches to discreetly discuss these matters.}'' 
Such clandestine conversations often revolve around academic help, with C02 pointing out that a specific chat group on their smartwatch, cheekily dubbed the ``\textit{Homework Help Squad},'' was dedicated to sharing answers. 
Compared to smartphones, smartwatches are less visible because of the smaller appearance and closer to the body.
The smartwatch's secretive nature makes it a preferred choice over phones, as C02 further added, ``\textit{Phones have passwords, but parents often know them. 
On checking, they would find our chat filled with homework answers, which is why we mostly use the smartwatch for such discussions.}''
This secrecy extends beyond academic discussions. 
C18 described how her friend and her boyfriend contacted each other through their smartwatches after an argument at school underscoring the device's role in navigating the complexities of teenage relationships. 
C02 also mentioned that in school, open conversations with a bullied classmate would typically be avoided, but on the smartwatch, their interactions were frequent. 
These instances illuminate that the smartwatch is not just a communication tool; it offers adolescents a new, more private space for socialization, aiding them in navigating intricate social terrains.\looseness=-1

\subsection{Socializing with Online Peers via Smartwatches}
\label{sec:online_peers}
Beyond interacting with local peers, many adolescents actively connect with online peers using their smartwatches, leveraging the device's built-in social features~(Section~\ref{sec:research_context}).
Within this context, smartwatches have given rise to a distinctive online phenomenon known as the ``表圈~(\textit{biaoquan})''. 
This virtual realm transcends the traditional messaging or notification capabilities of smartwatches and evolves into a vibrant social ecosystem~(Figure~\ref{fig:ecosystem}). 
Here, adolescents are not only seeking friendships but also chasing popularity and peer status.

As shown in Table~\ref{tab:concepts}\footnote{In this paper, we use \textit{pinyin} to refer the key concepts such as ``表圈~(\textit{biaoquan})'' and ``处关系(\textit{chu guanxi})''~\cite{yangVirtualGiftsGuanxi2011} as there is no appropriate English phrase to convey their specific meaning in Chinese culture context. Instead, we use likes and visibility to refer ``赞'' and ``热度'' to help readers connect to previous relevant work~\cite{scissorsWhatAttitudesBehaviors2016,nesi_transformation_2018-1,nesi_transformation_2018-2}.}, \textit{Biaoquan} refers to the social community within the smartwatch platform, where users, commonly referred to as ``圈内人~(\textit{quanneiren})'', actively seek likes and visibility which signify peer status. 
Notably, a user with over 200k likes and a high level of visibility is recognized as the ``大佬~(\textit{dalao})''. 
Adolescents in \textit{biaoquan} actively engage in ``处关系(\textit{chu guanxi})'', establish or join ``家族~(\textit{jiazu})'', and work on ``扩列~(\textit{kuolie})'', not just within the smartwatch ecosystem but also beyond, all in a bid to enhance their peer status~(Figure~\ref{fig:ecosystem}).
Remarkably, as users elevate their social status to become \textit{dalao}, they in turn enjoy enhanced privileges in those activities. 
This dynamic establishes a feedback loop, reinforcing their prominence and creating a closed circle within \textit{biaoquan} ecosystem.
In this section, we begin by introducing the social capital and activities within \textit{biaoquan}. 
Next, we discuss the privileges of \textit{dalao} and the nature of friendship in this online community.\looseness=-1

\subsubsection{Social Capital: Likes and Visibility as Peer Status}

Within \textit{biaoquan}, likes and visibility serve as social capital, signifying peer status.
Likes and visibility are the foundation of \textit{biaoquan}, serving as the threshold for various social activities.
``\textit{Likes are the universal currency within biaoquan. Likes can buy you praise, admiration, great popularity, and even thers give you a lot.~(C07)}''
Visibility is another metric in \textit{biaoquan}, introduced as the excessive number of likes by many \textit{quanneiren}.

\begin{table*}
    \caption{Concepts related to the smartwatch mediated socialization referred by our participants.}
    \label{tab:concepts}
    \begin{tabular}{llp{27pc}}
    \toprule
    \textbf{Term} & \textbf{Literal Meaning} & \textbf{Free Translation} \\
    \hline
    碰一碰 \textit{pengyipeng} & Bump-to-Connect & a feature of XTC smartwatches which allows adolescents to add each other to their friend lists by simply touching their watches together. \\
    微聊 \textit{weiliao} & MiniChat & the instant messaging App on XTC smartwatches. \\
    好友圈 \textit{haoyouquan} & Friend Circle & the social media platform on XTC smartwatches where users can post updates and photos for their friends to view, like, and comment on. \\
    表圈 \textit{biaoquan} & watch circle & the social community based on XTC smartwatches, where users pursue likes and visibility. \\
    赞 \textit{zan} & likes & likes given to an individual's profile, with a daily limit of 20 likes per user to gift. \\
    热度 \textit{redu} & visibility & the degree to which a user is known or recognized within the watch circle. \\
    圈内人 \textit{quanneiren}  & circle members & the recognized users within the watch circle, who are actively seeking likes and visibility. \\
    大佬 \textit{dalao} & big shot & a user with over 200k likes and a high level of visibility within watch circle, indicating a high peer status. \\
    双互 \textit{shuanghu} & mutual liking & the daily practice of reciprocally giving likes to each other. \\
    处关系 \textit{chu guanxi} & build relationships & the act of establishing virtual relationships between one adolescent and others, through the exchanges of favor and responsibilities. \\
    家族 \textit{jiazu} & house & a social group comprised of \textit{quanneiren} within watch circle, identified by a unified prefix in the members' usernames. \\
    扩列 \textit{kuolie} & friend list expansion & the act of seeking, making, and adding more friends to the contact lists of the built-in social networking platform \textit{MiniChat}. \\
    网通 \textit{wangtong} & online wanted & the act of exposing someone's misconduct to a broader audience through posters on smartwatch platforms with the intent to ruin someone's reputation.
    \\
    \bottomrule
    \end{tabular}
\end{table*}

\paragraph{\textbf{Likes}.}

\textit{Quanneiren} seek likes as ``\textit{essential as daily sustenance, engaging with them each day~(C14)}.'' 
Several interviewees emphasized that the ``\textit{pursuit of likes equates to self-pleasing and a sense of accomplishment~(C17)}'', and the publicness of likes on the profile akin to a ``\textit{luxury handbag with a transparent price tag~(C08).}''

In \textit{biaoquan}, likes are the most important measure of a person.
As C15 explained, ``\textit{If you have many likes on your profile, it means you have many friends. Many people are willing to like your posts, which implies you might be of better manners. So many people want to add you as a friend.}''
Consequently, the quantity of likes became a qualification for many social activities in \textit{biaoquan}.
C14 said, ``\textit{Without likes, no one cares about you, no one likes you, no one chats with you. I want to have more friends, so I have to collect likes.}''
The number of likes is a determinant of peer status in \textit{biaoquan}, and those with more likes are recognized as having higher status, \ie \textit{dalao}.\looseness=-1

To increase their likes, \textit{quanneiren} employed various tactics such as promises, harassment, coercion, and hacking. 
The practice where adolescents committed to liking each other's posts daily is termed \textit{shuanghu}. 
As C12 mentioned, ``\textit{Adding friends is just for shuanghu}.''
If promises fail to garner enough likes, some adolescents harassed their friends for them. 
For example, C02 and C16 confessed to sending many messages urging friends to like their profiles. 
C05 spoke of a friend who made over 30 calls seeking likes, and C14 was inundated with early morning messages demanding the same. 
A few even threatened to unfriend those who did not like their profiles~(\eg C03). 
Additionally, a few have cleverly exploited system bugs to amass likes quickly without any social engagement.

\paragraph{\textbf{Visibility}.}

When likes are high for everyone by accumulation, \textit{quanneiren} introduced visibility as a new social capital.
Unlike likes, smartwatches do not provide a follower function, rendering visibility invisible. 
It depends on how many \textit{quanneiren} know the user. \looseness=-1

Getting hundreds of thousands of likes was the first step to visibility~(\eg C13). 
Then, adolescents adopted several strategies within smartwatch. 
Some adolescents used photos and captions found online to impersonate themselves~(\eg C10, C13, C14, C15). 
Another method was to reciprocate actions, such as forming virtual relationships~(\textit{chu guanxi}, Section~\ref{sec:virtual_relationships}), and mentioning each other frequently in \textit{Friend Circle}~(\eg C10, C13, C14, C15).
Forming \textit{jiazu} was a key method to enhance visibility~(Section~\ref{sec:jiazu}). 
Boosting visibility by these ways demands time and persistence~(\eg C10). 
So, some adolescents amplified their visibility swiftly through controversies, including sparking societal debates or criticizing celebrities~(\eg C11, C15) and \textit{dalao}~(\eg C11, C12, C13, C14). 
Beyond smartwatches, adolescents increased their \textit{biaoquan} visibility on other social platforms like Xiaohongshu and Douyin, by creating personal tags, sharing content of smartwatch, and forming fan groups.
Overall, after gaining enough likes, teens gained more visibility in \textit{biaoquan} by crafting personas, reciprocating, creating controversies, sharing content, etc. inside and outside the smartwatch.

\correct{Many participants said they invested more time in social interactions on smartwatches compared to other social media platforms such as Xiaohongshu and Douyin~(\eg C07, C08, C12, C14, C15, C17, C18) cause smartwatches are always with them while smartphones as well as these mobile applications are limited by parental oversight. 
In addition, C18 shared that, on those mobile platforms, she consistently struggled to garner likes, as her content lacks of richness and visual presentation compared to adults.
However, on smartwatch platforms, she could accumulate more likes through socializing with other peers~(described in Section~\ref{sec:virtual_relationships}, Section~\ref{sec:jiazu}, and Section~\ref{sec:kuolie}). 
Reversely, the number of `likes' and the extent of visibility, as a form of social capital, signify the peer status and privilege~(described in Section~\ref{sec:dalao}) in participating smartwatch-based social activities~(e.g., \textit{chuguanxi}, establishing or joining \textit{jiazu}, and \textit{kuolie}). 
Such privilege stimulates the vanity of adolescents, prompting them to continue to engage in these activities to accumulate more social capital, creating a closed loop.
Smartwatches, with their easy-accessible, exclusive, peer-based social ecosystem, serve as the common ground for such closed-loop socialization. 
Therefore, the intricate relationship between social caption, social activities, and peer status in smartwatch-based socialization isolates likes and visibility from the ones on mobile platforms where likes typically represent superficial content popularity and the number of followers.
The following sections detail the social activities and the representation of peer status and privilege.
}


\subsubsection{Social Activity: \textit{Chu Guanxi} and Virtual Relationships}
\label{sec:virtual_relationships}
\begin{figure*}
 \includegraphics[width=\linewidth]{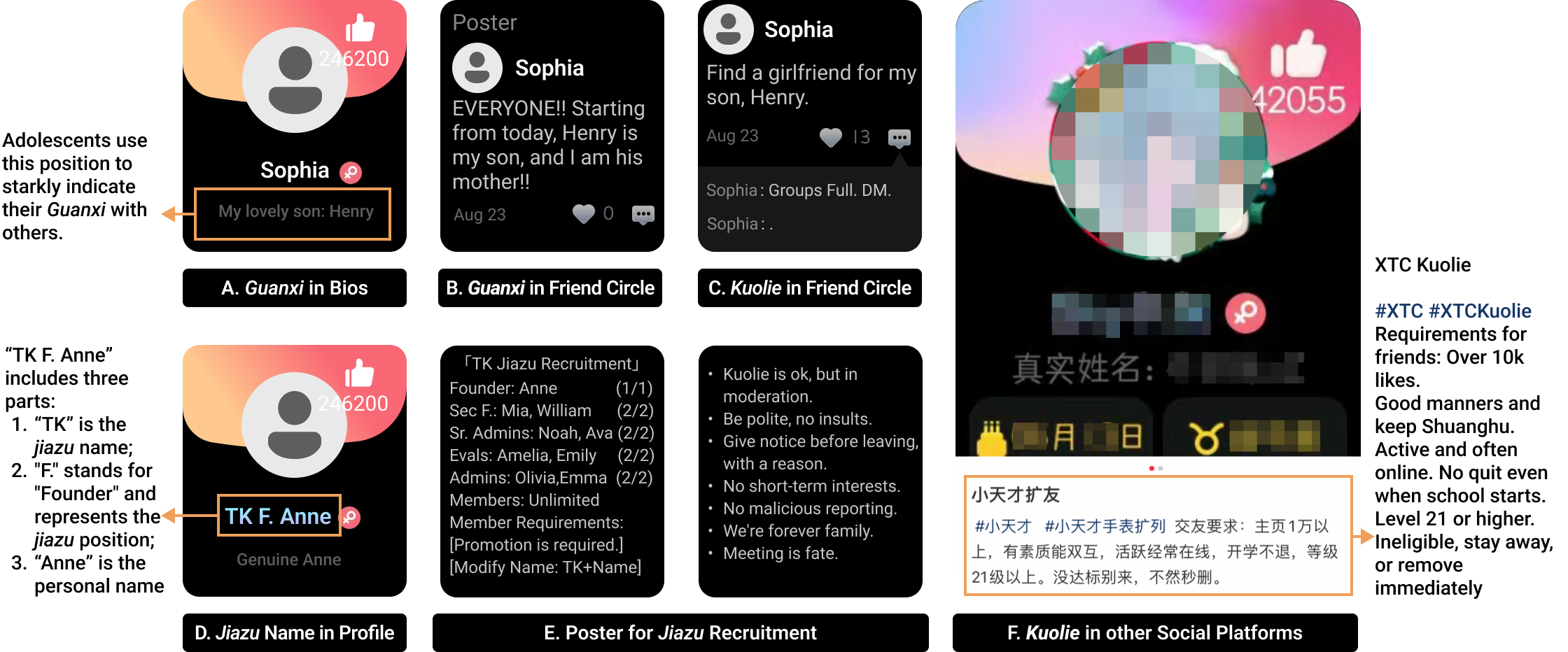}
 \caption{Examples of how adolescents engage in social activities on smartwatches. 
 A. Indicating \textit{Guanxi} with others in their bios. B. Posting \textit{Guanxi} established with others in their \textit{Friend Circle}. C. Seeking \textit{Kuolie} within the \textit{Friend Circle}. D. Displaying their \textit{Jiazu} and their positions within the family on their profiles. E. Posters recruiting for \textit{Jiazu}, including information about the \textit{Jiazu} and recruitment requirements. F. Engaging in \textit{Kuolie} activities on other social platforms. To safeguard the participants' privacy, we have redrawn Figures A-E based on the supplementary materials provided by them, using fictitious names and modified content.}
 \Description{Examples of how adolescents engage in social activities on smartwatches. 
 A. Indicating Guanxi with others in their bios. B. Posting Guanxi established with others in their Friend Circle. C. Seeking Kuolie within the Friend Circle. D. Displaying their Jiazu and their positions within the family on their profiles. E. Posters recruiting for Jiazu, including information about the Jiazu and recruitment requirements. F. Engaging in Kuolie activities on other social platforms. To safeguard the participants' privacy, we have redrawn Figures A-E based on the supplementary materials provided by them, using fictitious names and modified content.}
 \label{fig:social_activities}
\end{figure*}

Within \textit{biaoquan}, adolescents formed virtual relationships that mirror real-life interpersonal and familial connections, such as lovers, besties, mentors, disciples, and even parents and children, for the sake of improving peer status. 
This process usually unfolded in four stages.
First, initiators expressed interest in forming relationships, often using compliments and sometimes virtual points as gestures. 
For instance, C17 spent 30 points to make a best friend, while C18's admirer requested mentorship due to C18's popular posts. 
Once there was mutual agreement, participants made an ``\textit{official announcement}'' on platforms like \textit{Friend Circles}~(Figure~\ref{fig:social_activities}B) and their profile bios~(Figure~\ref{fig:social_activities}A). 
C18 observed an announcement where a girl introduced another user as her `son', with the latter acknowledging her as `mom'.
These relationships came with responsibilities. 
Authority figures, such as mentors or virtual parents, provided guidance and support to their subordinates in how to better live in \textit{biaoquan} like collecting  likes and building more relationships. 
For example, C16 taught his apprentice hacking techniques for more likes, and a virtual mother sought friends for her virtual son.
However, these relationships are often short-lived. 
C16's apprentice ended their mentorship after surpassing C16 in likes, and C17's point-bought friend was soon lost amidst numerous other friends.
In conclusion, the formation of virtual relationships among adolescents in \textit{biaoquan} followed a sequence of initiation, mutual agreement, public declaration, and evolving responsibilities. 
These relationships were transient in nature but often involve social capital flows between different peer status.

\subsubsection{Social Activity: Establishing or Joining in Jiazu}
\label{sec:jiazu}

\textit{Jiazu} is similar to ``house'' in \textit{biaoquan}.
Most adolescents established or joined \textit{jiazu} for more friends, likes and visibility by interacting with other members within the \textit{jiazu}.
There were structured hierarchies in \textit{biaoquan} \footnote{Ranging from high to low, these positions include Founder~(\ie patriarch, 1 person), Secondary Founders~(2-3 people), Senior Administrators and Evaluators~(more than 3 people), Administrators, Evaluators, Promoters~(more than 3 people), and Regular Members.}.
Higher position meant greater exposure.
For example, ``founder'' were prominently featured on promotional posters, which were widely circulated by fellow members.

Most activities within \textit{jiazu} happened in group chats, including playing text-based games~(\eg C07), sharing daily life~(\eg C08, C15, C16) with peer support~(\eg C05, C07). 
For instance, C07's friend shared personal issues such as parents' divorce or lack of parental attention, obtaining comfort from other members. 
In addition, \textit{jiazu} served as a ``\textit{backer}''~(\eg C07, C12, C05, C10, C08), providing support and advice on negotiation or counteractions when its members are in conflict with users outside the \textit{jiazu}.  
As described by C05, ``\textit{if someone offends them, the jiazu members unite to criticize them together.}'' 
In \textit{jiazu}, adolescents forged distant friendships, engaged in constructing and simulating social hierarchies, shaping collective identities and responsibilities while seeking collective refuge.
In general, \textit{jiazu} exhibited four characteristics.\looseness=-1

\paragraph{\textbf{Networking}.}
The development of \textit{jiazu} members was centered on the patriarch's social circle and spreads outward. 
This was evident from C03's description of how they built a \textit{jiazu}: ``\textit{At first, I invited a close friend of mine, and then we invited those we have good relationships with, and they, in turn, brought in their own well-connected friends.}''

\paragraph{\textbf{Membership}.}
Joining a \textit{jiazu} required meeting certain conditions. 
These ranged from hard conditions, like accumulating a significant number of likes, to softer criteria, such as having a good personality and avoiding disputes~(Figure~\ref{fig:social_activities}E). 
Once inducted, members added a unified prefix, often the \textit{jiazu} name, to their usernames to signify their membership~(\eg C03, C07，C16). 
For example, the name \textit{``TK F. Anne''} combined a \textit{jiazu} name~(``TK''), the position in \textit{jiazu}~(Founder) and a personal name~(``Anne'')~(Figure~\ref{fig:social_activities}D).
This served as a badge of honor and association. 
Therefore, users with lower status often seeked to join well-known \textit{jiazu} to improve their visibility and peer status. 

\paragraph{\textbf{Spontaneity}.}

In fact, smartwatches do not feature any specific functionalities related to \textit{jiazu}.
Teenage users have extensively repurposed existing features, such as group chats and username, to fulfill various activities of \textit{jiazu}.
This adaptability posed significant challenges in terms of \textit{jiazu} management and upkeep. 
For instance, as explained by C07, group chats on smartwatches were limited to 11 participants. 
Due to this limitation, \textit{jiazu} often became quiet: ``\textit{in the first few days, everyone had something to share, but later, there was nothing left to discuss.~(C07)}''
The group chat was only for chatting, and the lack of other features also makes \textit{jiazu} very difficult to manage~(\eg C05, C08), C05 said ``\textit{actually, I couldn't do anything and I couldn't ban anyone, so I finally had to disband the group.}''
Beyond disbanding \textit{jiazu}, they lack effective measures like muting or other moderation actions~(\eg C08, C10, C15).

\paragraph{\textbf{Ephemerality}.} 
As noted by C07 and C11, most \textit{jiazu} lasted only a few days or a few weeks, with the longest survival being around two months. 
\textit{Jiazu} dissolved for reasons ranging from inactivity~(\eg C07, C11, C15), managerial complexity and conflict~(\eg C07, C11). 
Adolescents initially established or join \textit{jiazu} primarily to meet new people. 
Upon joining, they swiftly engaged in sharing interesting topics, forming private connections, and expanding their networks within a few days. 
Subsequently, these \textit{jiazu} either became quiet or riddled with repetitive conflicts.
As C07 remarked, ``\textit{Jiazu tended to quiet quickly, then many people left the jiazu because there was no one chatting in the group.}''
C03 mentioned that a prior reason for leaving a \textit{jiazu} was constant, tedious arguments.
Following the dissolution of a \textit{jiazu}, adolescents proceed to establish or join new families to once again expand their social circles. 
Also some found these \textit{jiazu} uninteresting and choose not to participate further.
Moreover, defeating after flame war with competitors was an important reason of \textit{jiazu} dissolution~(\eg C03, C12), as discussed in Section~\ref{sec:flame_war}.

\subsubsection{Social Activity: Kuolie to Expand Social Circles}
\label{sec:kuolie}
Adolescents engaged in \textit{kuolie}, which involves actively seeking and adding friends, to increase their chances of receiving likes and enhancing their visibility. 
They expanded their friend list through various channels, such as group chats, \textit{jiazu}~(Section~\ref{sec:jiazu}), and other social platforms like Xiaohongshu. 
In particular, adolescents ``cast a wide net'' by post their details and criteria on these platforms, awaiting contacts~(Figure~\ref{fig:social_activities}F).
These criteria included both ``hard'' ones like user levels~(typically set at 20 for 20 daily likes) and likes count~(indicates one's \textit{biaoquan} activity). 
``Soft'' criteria included shared interests~(\eg C16, C17) and good manners, with a preference against friends who show off, cheat, swear, or spread negativity~(\eg C12, C13, C16). 
Beyond these direct efforts, some adolescents also utilized intermediaries to conduct \textit{kuolie}.
As mentioned by C17, these intermediaries introduced the individual on \textit{Friend Circle} by showcasing specific criteria that the seeking individual desires in potential friends~(Figure~\ref{fig:social_activities}C). 

\subsubsection{Social Prestige: Being a \textit{Dalao}}
\label{sec:dalao}
As previously mentioned, \textit{likes} and \textit{visibility} constituted the peer status within \textit{biaoquan}, with adolescents possessing this peer status being referred to as \textit{dalao}. 
\textit{Quanneiren} were proud to be friends with \textit{dalao} when expanding their friend lists. 
Therefore, \textit{dalao} enjoyed privileges across various social activities within \textit{biaoquan}.
\textit{Dalao} reaped the benefits without effort when improving likes and increasing. 
As noted by C07, ``\textit{Many people actively and voluntarily give likes to dalao.}'' 
When establishing virtual relationships, \textit{quanneiren} were flocking to \textit{dalao}, as described by C12, ``\textit{Many people want to have relationship with dalao; whether it happens or not depends on dalao's mood.}'' 
C12 explained that someone she knew who had the \textit{dalao} title was highly sought after: ``\textit{Everyone wants to add him.}'' 
These privileges of \textit{dalao} extended to the \textit{jiazu} they create. 
\textit{jiazu} founded by \textit{dalao} received more attention than other \textit{jiazu}, and many individuals wished to join \textit{jiazu} created by \textit{dalao} to ride \textit{dalao}'s popularity~(\eg C07, C14). 
Consequently, joining \textit{jiazu} created by \textit{dalao} came with higher criteria~(\eg C10).
However, the title of \textit{dalao} in \textit{biaoquan} was a double-edged sword due to more attention and higher expectations than the average user.
For instance, C15 said, ``\textit{I can't just delete my friends. Otherwise, they might say I'm showing off, which could impact my reputation.}''\looseness=-1

\subsubsection{Friendships within Biaoquan}
In \textit{biaoquan}, the majority of friendships evolved rapidly and endure only briefly. 
As expressed by C11, ``\textit{You can meet someone one minute, become close friends in three, have endless conversations in five, and then stop talking after ten.}'' 
Due to the limitations of smartwatches on the number of friends, most adolescents swiftly deleted friends who fail to meet their expectations, such as liking their posts or engaging in active chats~(\eg C13, C14, C15, C17). 
This pattern of swift development and dissolution led adolescents to perceive most \textit{biaoquan} friendships as fragile and insincere. 
As C13 said, ``\textit{Around 80\% of these friends are fake... I don't want to keep these friends because they are too insincere... Those fake friends are only here to follow trends and boost their own visibility...}''  
C15 added, ``\textit{Some people pretend to be friends with you for the sake of your visibility. Some genuinely want to be real friends. But in a short time, it's hard to tell whether they are fake or real.}''
Nonetheless, a few adolescents did gain genuine friendships within \textit{biaoquan}. 
As C15 pointed out, ``\textit{I've made some good friends through smartwatch. I can share life's pressures with them and discuss things I can't talk about in real life. We vent together and also share interesting things.}''
When they aspired to transform \textit{biaoquan} friendships into enduring, stable relationships, adolescents often sought to establish connections on other social platforms such as QQ and WeChat, as shown in Figure~\ref{fig:ecosystem}. 
This allowed them to distinguish these bonds from the predominantly superficial and transient friendships within \textit{biaoquan}~(\eg C09, C10, C15, C18).

\subsection{Negative Social Dynamics within Smartwatches}
\label{sec:negative}
In smartwatches, negative social activities folloedw established procedures, such as \textit{drama}, \textit{wangtong}, and \textit{flame war}. 
These activities were driven by concerns about personal reputation and collective honor than by material interests.

\subsubsection{Discrimination in and outside the biaoquan}

Discrimination happened in both offline and online settings.
For example, C06's experience underscored a tangible discrimination based on the brand and authenticity of watches, revealing that some teenagers face ridicule because they may have bought ``\textit{a kind of fake smartwatch}'' and when peers tried to connect with them, they found out they ``\textit{can't add them, and then they mock them.}''
Moreover, the likes on smartwatches have emerged as a new source of online social discrimination. 
Some participants expressed the reality where those with more likes might criticize others, stating that there is a bias against those with fewer likes on their profile, and some even think, ``\textit{How can they join \textit{biaoquan} with so few~(likes)?}''~(\eg C11, C14) or more directly, ``\textit{those with more likes discriminate against those with fewer likes}''~(\eg C06, C08).
Moreover, most \textit{quanneiren} had a sense of superiority, leading to a stark division between insiders and outsiders, with C11 indicating that if you do not join, you might be perceived as ``\textit{not keeping up with the times or trends, and then they will inexplicably scold you.}'' \looseness=-1

\subsubsection{Drama to catch eyes}

Within the smartwatch social ecosystem, adolescents often sought \textit{visibility} through drama. 
Many befriended prolifically, seeking immediate recognition, only to subsequently cut ties; some even lashed out without discernible reason, only to later play the reformed character.
Such behaviors deliberately incited conflicts, within which certained teenagers adopt a performative victimhood stance to dodge accountability and seeking sympathy for their actions. 
C05 detailed tactics of playing such victim: sharing selective screenshots, manipulating narratives, and gaining sympathy.
As described by C05, ``\textit{Some use a very pitiful tone... post a screenshot, but only the one where someone is scolding them...}''
By spotlighting only the accusations they face, they intentionally bypassed their potential provocations, curating a skewed representation for eliciting sympathy.
Alarmingly, as C12 and C15 pointed out, there were those who feign mental distress to further their visibility, leading peers to unknowingly support their ruse. 
Additionally, C11 pointed out the inflation of trivial squabbles, ``\textit{... he'd ask his friends to share the matter, turning a minor issue into something everyone talks about...}''
Such instances amplified insignificant disagreements, roping in a larger audience, only to see the main actors reconcile, leaving onlookers baffled while increasing their visibility.\looseness=-1

\subsubsection{Wangtong to ruin someone's reputation by Online Wanted}

Within \textit{biaoquan}, \textit{wangtong} is a unique form of cyberbullying, which translates to ``online wanted.'' 
Distinct from conventional online wanted posters, \textit{wangtong} in \textit{biaoquan} tended to focus on publicizing an individual's alleged wrongdoings to a wider audience, tarnishing their reputation, leading to unfriending, and preventing further social connections within \textit{biaoquan}~(\eg C13, C5), rather than locating the individual.
The genesis of \textit{wangtong} typically arised from conflicts between two \textit{quanneiren}, such as verbal insults or offensive remarks towards one another or even the nation.
The general process of \textit{wangtong} unfolded as follows: following a conflict, teens create simple posters on their smartwatches to showcase the alleged wrongdoings of the other party~(\eg C12). 
Subsequently, they and their friends shared these posters on their \textit{Friend Circle}and encouraged others to repost~(\eg C15).
Several factors contributed to the emergence of \textit{wangtong}. 
Firstly, \textit{biaoquan} had a relatively small user base and a fairly uniform user demographic, enabling such public condemnation to attain high visibility. 
Secondly, one's social standing and reputation were integral aspects of visibility within \textit{biaoquan}. 
Public shaming, as observed in \textit{wangtong}, served as a form of deterrence.
However, it is worth noting that \textit{wangtong} often increasing the target's visibility. 
Many adolescents intentionally subjected themselves to \textit{wangtong} to garner attention.\looseness=-1

\subsubsection{Flame War for the honor of jiazu}
\label{sec:flame_war}
Flame War refers to conflicts within \textit{jiazu} in \textit{biaoquan}. 
Unlike traditional online hostility and verbal abuse, flame wars typically involved multiple members from two or more \textit{jiazu} who engaged in these conflicts not for personal reasons but on behalf of their respective \textit{jiazu}~(\eg C10, C12).
The origins of a flame war between \textit{jiazu} usually stemmed from instances where a member of one \textit{jiazu} experienced insults or mistreatment. 
In such cases, the victim gathered evidence and presented it within their \textit{jiazu}~(\eg C12). 
Subsequently, other members of that \textit{jiazu} rallied to support the victim by resorting to heated arguments and demanding apologies~(\eg C05, C08, C12).
Because the reliance provided by \textit{jiazu} was a very important reason why people choose to join jiazu, if they could not help the members to get justice, then the members would push out \textit{jiazu} and the reputation of \textit{jiazu} and its founder would deteriorate rapidly~(\eg C10).
If this approach failed to resolve the issue, the founder of the \textit{jiazu} escalated the conflict into a full-fledged war involving both the victim's \textit{jiazu} and the aggressor's \textit{jiazu}~(\eg C10).
These wars often involved members from at least two \textit{jiazu}, and sometimes even a third \textit{jiazu} if they had a bond relations with either of the two conflicting \textit{jiazu}~(\eg C10). 
The disputes in these wars no longer revolved around specific issues but were driven by a sense of \textit{jiazu} pride and honor. 
Because the losing \textit{jiazu} face ridicule from the victors and others, causing a loss of cohesion and leading to the dissolution of \textit{jiazu}~(\eg C03).\looseness=-1

\subsection{Perceived Benefits and Challenges of Smartwatch-based Socialization}
\label{sec:benefitsandchallenges}

Smartwatches rely on their physical affordances and always-on-wrist presence, providing adolescents a convenient communication channel.
Also, the exclusivity of the built-in social platforms fostered a safe environment to build connections.
Nonetheless, current design remained immature and even contained some manipulated features and lacked effective moderation in peer-to-peer communications.
In this section, we reported the perceived benefits and challenges of smartwatch-mediated socialization, providing insight into future design.\looseness=-1

\subsubsection{Benefits}
\label{sec:benefits}
We identified three perceived benefits of smartwatch-based socialization, including:

\begin{itemize}
    \item \textbf{Portability and Convenience of Smartwatches.} Participants often highlighted the portability and convenience of smartwatches. The voice input feature enabled users to communicate effortlessly, free from the constraints of typing. As C04 stated, \textit{I really enjoy chatting with my friends, but typing can be cumbersome. With the smartwatch, I can simply speak directly.}'' Furthermore, the physical attributes, such as their lightweight design and compact size, made smartwatches unobtrusive. Their always-on-wrist presence ensured they are \textit{always within reach}''~(C06). C05 remarked, \textit{I often prefer not to take out my phone because it's somewhat heavy. Thus, I engage in chats more frequently on the smartwatch.}'' Additionally, the integration of intuitive gestures, like swipes or taps, provided quick and effortless access to connect with peers. C06 noted, ``\textit{With this watch, you swipe up, left, or right, and you can access MiniChat and view the messages.}'' As a result, the majority of adolescents expressed an increased willingness to chat, share, and engage on the built-in social platforms of smartwatches.
    \item \textbf{Exclusion of Predators in a Closed-type Social Environment.} Smartwatches operated in a ``\textit{closed system}''~(C07), offering a safer and more enclosed environment compared to other popular online platforms in China like QQ and WeChat. The limited exposure to the vast internet populace potentially reduced encounters with ill-intentioned individuals. In addition, the social platform within smartwatches was primarily populated by their peers, which allowed adolescents to share content freely without the fear of judgment or misinterpretation from older users. C10 likened it to a ``\textit{miniaturized society},'' but one that was more harmonious since the majority of users are elementary or middle school students. C08 shared that this exclusivity nature reassures their parents about the safety of their online interactions as it offered a ``\textit{closed-type social environment}.''
    \item \textbf{Smartwatches as Personal Sanctuaries Shielded from Parental Oversight.} Many participants saw smartwatches as personal sanctuaries, affording them a unique space that is less exposed to parental oversight compared to smartphones. For example, parents were less inclined to scrutinize smartwatches, with C02 noting ``\textit{Parents generally don't check smartwatches.}'' This reduced scrutiny aligned with a broader sentiment that smartwatches were viewed as intimate devices, evoking a strong sense of autonomy and individual ownership, as C11 recounted, ``\textit{My parents used to like looking at the chat history on my phone... but now they feel that the smartwatch is my personal item, and I don't really like sharing it with them.}''
\end{itemize}

\subsubsection{Challenges}
\label{sec:challenges}
We uncovered three representative challenges that bothered adolescents in socialization, including:

\begin{itemize}
    \item \textbf{Manipulative Design of Social features.} Manipulative design, \aka dark patterns~\cite{grayDarkPatternsSide2018}, in the smartwatch platform appeared to have a negative impact on the social behaviors of teenagers. C04 noted, ``\textit{I delete them and then add them back so that I can get points.}'' Such behaviors revealed how genuine social interaction is being overshadowed by the lure of rewards. C10 added, ``\textit{If you didn't 'like' my profile today, I might delete you tomorrow because I don't have enough friend slots}'', highlighting the design constraints can lead to impulsive decisions. Furthermore, certain social features were reserved for more advanced or newer models, creating a hierarchy and hindering peers to communicate as C03 mentioned, ``\textit{Even if they have Friend Circle, they can't see my post. It's disappointing cause they didn't know what I was talking about at all.}''
    \item \textbf{Non-Interoperability Between Brands.} We also found the frustrations arising from non-interoperability between different smartwatch brands. Users felt constrained, unable to connect with peers who had a different brand, as C01 said, ``\textit{Two people with different watch brands can't become friends, which is disgusting.}'' This limited their social circle and led to feelings of exclusion. C05 expressed a desire for smartwatches to function like mobile phones where different brands can easily connect, ``\textit{Chatting apps or third-party apps, like WeChat or QQ. Not everyone uses the same brand.}'' The prevalence of one brand over another in certain social circles also pushed individuals towards purchasing decisions based on peer pressure rather than personal preference.
    \item \textbf{Lack of Moderation in Communication.} Many participants reported experiencing verbal victimization, including verbal abuse and harassment. Despite the presence of official text content regulation, it seemed to have limited effectiveness. Some victims often found other means to bypass these restrictions, such as resorting to abbreviations and images. Additionally, while users could report uncomfortable content they see in \textit{Friend Circle}, reporting could be abusive. As C14 said, ``\textit{One of the most annoying things is that no matter what content I post, one person will just always report me and then I get banned or something like that.}'' 
\end{itemize}
\section{Discussion}
This study is the first to investigate into a comprehensive smartwatch-mediated social ecosystem for adolescents. 
Smartwatches have enhanced offline socialization among adolescents and spawned a variety of online social activities such as \textit{kuolie}, \textit{jiazu}, and \textit{chuguanxi}
In this section, we reflect on our findings and situate them within existing literature on adolescent online socialization. 
We then discuss how the unique affordances and Chinese culture influence adolescents use smartwatch for socialization, and explore how these insights can inform future research and design within the HCI and CSCW community.

\subsection{Smartwatch as a Social Platform for Adolescents}
Our research delves into Chinese adolescents' smartwatch-mediated offline and online social activities, expanding on prior studies centered on computers or smartphones~\cite{nesi_transformation_2018-2,buhler_how_2013,rusak_properties_2014,ali_understanding_2022}.
Overall, our findings align with the existing literature on adolescent online socialization.
Many social activities within smartwatches align with those depicted in prior literature, such as present themselves~\cite{yarosh_youthtube_2016}, make more friends~\cite{zywica_faces_2008}, and compensate for social phobias in real life~\cite{wang_mediator_2011} through online socialization.
Quantifiable metrics also regulate adolescents' smartwatch-mediated socialization~\cite{chua_follow_2016}, with likes serving as a primary driver of engagement in \textit{biaoquan}. 
Many adolescents aspire to connect with \textit{dalao}, aligning with prior research indicating that they enhance popularity by showcasing associations with high-status peers~\cite{marwick_i_2011}.
Smartwatches also make adolescents feel ``tethered''~\cite{fox_dark_2015}, particularly evident among \textit{dalao} in \textit{biaoquan}.
The negative impacts of online socialization on adolescents, as documented in prior research, are also apparent in smartwatch-based interactions, including cyberbullying such as sexting~\cite{razi_lets_2020,hartikainen_if_2021}, harassment~\cite{10.1145/3555136}, and drama~\cite{marwick_its_2014}. 
Meanwhile, some findings complement prior research, reflected in the unique affordances of smartwatches and possibly cultural differences among Chinese users, which will be further discussed in Section~\ref{sec:affordances} and Section~\ref{sec:cultural_factors}.\looseness=-1

\subsection{Physicality,  Locality, and Exclusivity: Leveraging Smartwatch Affordances in Peer Socialization}
\label{sec:affordances}
Smartwatch enhances adolescents' convenience and privacy through their physicality.
It starts their socialization from offline interactions, and creating an exclusive space for teenagers, distinguishing smartwatch from traditional online socialization via computers and smartphones.

\subsubsection{Physicality}
It is crucial to acknowledge the impact of the physical presence of smartwatches, constantly worn on the wrist, on adolescent social behavior.
The ``always on'' nature of these devices introduces a unique dimension to their social interactions, blurring the lines between online and offline engagement. 
The physicality of smartwatches fosters a sense of immediacy and accessibility, allowing adolescents to engage with their peers in a more spontaneous and continuous manner~\cite{cecchinatoAlwaysLineUser2017}.
They act as catalysts for face-to-face encounters, reflected in the \textit{Bump-to-Connect} feature, simplifying friend additions and promoting spontaneous interactions. 
Furthermore, smartwatches' physicality enhances adolescents' sense of ownership and privacy. 
Worn constantly, they become personal items. 
Many interviewees noted their parents paid little attention to their smartwatch use, granting them autonomy in online interactions. 
This heightened sense of ownership and privacy influences adolescents' social behaviors and comfort with these devices.\looseness=-1

\subsubsection{Locality}
In our study, we observed that smartwatch-based socialization is motivated by and rooted in the local context.
Smartwatches enable communication with offline friends, bridging distance and time to reinforce local social ties. 
Teens use \textit{Bump-to-Connect} to transition friendships from offline to online. 
Many participants use smartwatches for private discussions, particularly on school-sensitive topics like class gossip or sharing homework answers.
Typically, their smartwatch interactions begin with close offline contacts, such as classmates, during activities like \textit{kuolie} or \textit{jiazu}. 
These findings underscore the intricate relationship between smartwatch-mediated socialization and the local context, revealing the motivations, practices, and complexities that adolescents navigate in managing their digital and physical social realms.
\looseness=-1

\subsubsection{Exclusivity}

Access to the social ecosystem within smartwatches is contingent on owning a smartwatch.
This enclosed environment provides adolescents with a sense of security by excluding potential threats~(\ie adult predators without one kid smartwatch). 
For example, several participants expressed feeling more comfortable posting content on smartwatches than on other adult-dominant social platforms, like Xiaohongshu and Douyin, because smartwatches primarily host peers.
This insular environment often shields adolescents' social activities from adult oversight, but also leads to unchecked peer victimization. 
Unless teens report incidents, adults remain unaware of the harm.\looseness=-1

Exclusivity is also reflected in the dominant user age group of 12 to 15-year-olds. 
Before 12, adolescents primarily use smartwatches to communicate with offline friends. 
As they age, their focus shifts to online interactions, investing more in these relationships~\cite{hayPeerRelationsChildhood2004a,rubinPeerRelationshipsChildhood2011}. 
This shift drives them to seek more online connections and diversify their relationships. 
Since most Chinese teenagers have restricted access to smartphones or computers, they rely on smartwatches for socialization, leading to diverse social activities within the smartwatch ecosystem. 
However, by age 16, academic and high school responsibilities reduce their interest in smartwatch-based social activities, which many view as immature. 
This trend mirrors that of other social media platforms~\cite{arjan_age_2008, rusak_properties_2014,schoenebeck_playful_2016}.\looseness=-1

\subsection{Cultural Practices: Shaping Adolescent Smartwatch-based Socialization in Chinese Contexts}
\label{sec:cultural_factors}

In this paper, we explore how the social behaviors of Chinese adolescents on smartwatch platforms are influenced by~(1) the Chinese concept of \textit{guanxi} and~(2) collectivist cultural values.
\looseness=-1

First, research has shown that in China, Internet users exchange virtual gifts or currency to enhance \textit{guanxi}, denoting interpersonal influence and connectedness~\cite{yangVirtualGiftsGuanxi2011,wangHumanCurrencyInteractionLearning2008}. 
Such exchanges manifest in adolescents' ``\textit{shuanghu}'' behavior within \textit{biaoquan}, whereby they commit to reciprocally giving profile likes. 
Beyond just exchanging likes, they deepen connections through \textit{chu guanxi}, forming virtual relationships such as lovers, mentors, or even parental figures. 
In these virtual spaces, teenagers can explore and experiment with different social roles without the societal pressures they face offline. 
This mirrors their experimentation with roles like marriage in gaming~\cite{freeman_simulating_2015}.
Meanwhile, such virtual relationships entail responsibilities for both parties, encapsulating a unique dynamic of obligation and indebtedness~\cite{yangGiftsFavorsBanquets2016,hwangFaceFavorChinese1987} characteristic of the \textit{guanxi} practice in Chinese society.
\looseness=-1

Second, influenced by the collectivistic culture grounded in Confucianism, Chinese social media users frequently seek belongingness and shape their identities through memberships and relationships~\cite{wangWeChatMomentsInternational2023}. 
This inclination is evident in the adolescents' \textit{jiazu} phenomenon. 
Members of a \textit{jiazu} adopt the same username prefix, symbolizing membership and fostering a sense of belonging. 
In China, the \textit{jiazu} concept surpasses mere familial ties, extending to a broader spectrum of social and virtual communities. 
Within these communities, individuals consistently uphold various obligations and responsibilities towards others, establishing the foundation of their societal interactions~\cite{heineThereUniversalNeed1999}. 
Such intimate connections not only encourage mutual support within the \textit{jiazu} but also spark flame wars to defend their \textit{jiazu}'s honor.\looseness=-1

In conclusion, this paper accentuates the deep-rooted influence of Chinese cultural aspects on adolescent social behaviors within smartwatch platforms. 
As technology continually intertwines with culture, it becomes paramount to comprehend the cultural subtleties steering behaviors in emerging digital domains. 
Future endeavors should further dissect how cultural norms mold behaviors across diverse digital platforms and various cultural contexts.\looseness=-1

\subsection{Design Implications}

\subsubsection{Design for Intentional Smartwatch Use}
In Section~\ref{sec:challenges}, we highlighted the presence of dark patterns~\cite{grayDarkPatternsSide2018} in smartwatches. 
\correct{Compared to other social media platforms, dark patterns in XTC also aim to engage users for extended periods and encourage social networking~\cite{mildner_about_2023}. Additionally, XTC's dark pattern pushes users to buy the latest version of the smartwatch by introducing enticing new features.}
These designs negatively influence teenage social behaviors, encouraging inauthentic interactions for rewards, impulsive decisions due to design constraints, and creating communication hierarchies based on device models.
Therefore, we advocate for ethical design, which includes reducing dark patterns, especially those that induce social anxiety in adolescents. 
It's essential to provide users with customizable control options, allowing them to manage their social experiences autonomously. 
Additionally, we recommend offering features that monitor and remind users of their screen time. 
This will help them maintain a balance between their virtual and real lives, encompassing tools for time management and rest reminders~\cite{hinikerPlanPlaySupporting2017,hinikerCocoVideosEmpirical2018}.\looseness=-1

\subsubsection{Design for Enhanced Social Connection}
Currently, XTC smartwatch users cannot connect with users of other smartwatch brands, hindering both offline and online social interactions for many teenagers. 
A significant number of participants have expressed a desire for cross-brand connectivity. 
We urge brands to collaborate and break down these barriers, fostering a more connective ecosystem for adolescent smartwatch users. 
Additionally, the use of smartwatches by teenagers comes with several restrictions: restricted character counts for posts in \textit{Friend Circle}, mandatory watermarks on screenshots, limited app choices, and constrained connectivity with other social media platforms and devices, which limit adolescent expression. 
Moreover, we advocate that future researchers can provide adolescents with more expression and communication channels based on smartwatches, such as using biosignal to enhance social connections~\cite{liuAnimoSharingBiosignals2019}.

\subsubsection{Design for Secure Social Ecosystem}
In Section~\ref{sec:challenges}, we highlighted that users often bypass smartwatch text regulations using abbreviations and images, leading to the spread of offensive language. 
This contributes significantly to cyberbullying within \textit{biaoquan}. 
To counter this, smartwatches have introduced sensitive reporting systems that suspend reported accounts swiftly. 
Yet, the misuse of these mechanisms has exacerbated cyberbullying. 
We suggest developers to build machine learning models for content moderation to detect harmful language and malicious reports, even in the form of abbreviations, aiming to foster a responsible social environment. 
Also, while pursuing \textit{kuolie}, many teenagers publicly share personal details, including phone numbers and birth dates, on social media. 
Future designers could use persuasive design~\cite{Persuasive_Design} to educate adolescents about online privacy risks and ways to safeguard their personal data during online interactions.\looseness=-1


\subsection{Limitations and Future Work}

This work has the following limitations:
(1) All participants are users of XTC smartwatches, as this brand is the most popular in China and we chose this representative platform as a case study for the topic. 
However, it is worth noting that there are several other smartwatch brands in China, such as HUAWEI, XIAOMI, and OPPO. 
Users of these alternative platforms might experience social exclusion by the mainstream users of XTC, as suggested by the interviewees in this paper;
(2) Although our research primarily focused on adolescents, parents can also play a significant role in smartwatch-mediated socialization. 
While parents may not oversee the specific content of communications on the smartwatch, XTC does offer parental features in their companion mobile applications, such as screen time oversight and app usage control;
(3) Like many interview-based studies, our research lacks a large-scale quantitative analysis. 
Qualitative analysis could provide a deeper understanding of smartwatch-mediated socialization, such as uncovering correlations between self-presentation/disclosure features and social status.

For future work, we suggest incorporating the perspectives of other stakeholders~(e.g., users of different platforms, parents, teachers, and manufacturers). 
After gaining insights into this unexplored phenomenon through interview-based qualitative research,
it would be highly advantageous to employ quantitative analyses.
Quantitative research can explore gender-based or age-based patterns in smartwatch usage, such as usage time, social behaviors, features engagement, as well as perspectives and influences among adolescents.
It would also offer potential disparities to consider regional differences.
Ethical issues, including data privacy and security, and possible psychological impacts on participants, should be carefully noted.
These quantitative inquiries aiming to a comprehension of usage behaviors could contribute to the refinement of smartwatch designs tailored to the unique social needs of adolescents.
We also suggest future research aimed at modeling socialization dynamics, present proof-of-concept interfaces for comprehensive design implication evaluations, explore various moderation strategies to curtail negative social activities, and develop more smartwatch-based social tools that help adolescents navigate peer experiences responsibly and safely.

\section{Conclusion}
We interviewed 18 Chinese adolescents~(aged 11-16) to explore their smartwatch usage for both offline and online social interactions. 
Our findings revealed a comprehensive smartwatch-mediated social ecosystem. 
These devices foster offline social bonds, enhancing private communication and easing initial interactions.
Online, various social activities on smartwatches are driven by likes and visibility, giving rise to diverse pursuits like \textit{jiazu}, \textit{kuolie}, \textit{chuguanxi}, and negative social dynamics. 
The unique affordances of smartwatches---physicality, locality, and exclusivity---set their social activities apart from other online socializing on computers or smartphones.
Eastern cultural values like \textit{guanxi} and collectivism further shaped these interactions, as adolescents navigated societal norms.
Our findings inform design implications for future wearables for younger populations, emphasizing safe and positive adolescent interactions.
\begin{acks}
This research was partially supported by the 2021 CCF-TencentRhino-Bird Research Fund and the Research Matching Grant Scheme(RMGS, Project No.9229095). 
We thank Tencent, China Computer Federation (CCF), and Research Grants Council of Hong Kong fortheir support and guidance.
We thank the teenagers who participated in our study.

\end{acks}

\balance
\bibliographystyle{ACM-Reference-Format}
\bibliography{reference}

\appendix
\section{Interview Questions}
\label{appendix:interview_questions}

\subsection{Motivation and Overall Usage}

\subsubsection{Why buy?}
\begin{itemize}
\item Could you tell us a bit about your smartwatch? (Show and Tell)
\item What made you decide to get this smartwatch in the first place?
\item Are you still happy with your smartwatch?
\item Do you ever think about getting a new one? If so, why?
\item If you had the chance, would you consider buying a different brand or model?
\end{itemize}

\subsubsection{Overall Usage}
\begin{itemize}
\item What do you usually do with your smartwatch? (Show and Tell)
\end{itemize}

\subsubsection{Self-Presentation \& Self-Disclosure within Smartwatch}
\begin{itemize}
\item Do you use the friend circle on your smartwatch?
\item Can you give us an idea of the kind of content you share? (Show and tell)
\item Do you pay attention to the number of likes and comments after posting?
\item Do you find smartwatch friend circles fun or boring?
\item Do you find your smartwatch friend circle convenient or inconvenient?
\item Do you use other platforms like QQ (Qzone), WeChat, Xiaohongshu, or Douyin (TikTok)?
\item Do you post on these platforms as well?
\item How is the content you post on other platforms different from what you post on your smartwatch?
\item Which platform do you prefer? Why?
\end{itemize}

\subsection{Online Socialization through Smartwatches}

\subsubsection{Socialization Online}
\begin{itemize}
\item Can you share an experience of making friends with someone using a smartwatch?
\item Do you have any requirements for adding friends?
\item Can you share an experience of adding friends using the "Bump-to-Connect" feature?
\item What do you think is different about "Bump-to-Connect" compared to other methods?
\item After adding friends, what do you typically do together?
\item Do you have any expectations or requirements for your friends?
\item Have you ever unfriended someone or been unfriended by someone?
\item Did you meet these friends in real life or only through the smartwatch?
\item Do you have other ways to stay in touch with them?
\item Is there any difference between communicating with these friends on the smartwatch and using QQ or WeChat?
\item Have you ever had any unpleasant experiences while communicating with these friends on the smartwatch?
\end{itemize}

\subsubsection{Online Communities \& Group Dynamics within Smartwatches}
\begin{itemize}
\item Do you pay attention to any specific data on your smartwatches?
\item How do you feel about the likes or thumbs-up feature on smartwatches?
\item What do you typically do on your smartwatches?
\item Do you know ``biaoquan''?
\item Do you know ``jiazu''?
\item Besides that, what other social activities do people do on their smartwatches?
\end{itemize}

\subsection{3. Offline Socialization around the Smartwatches}

\subsubsection{Socialization Offline}
\begin{itemize}
\item How many of your classmates or friends around you have smartwatches?
\item Do you and your classmates or friends talk about smartwatches?
\item Do you think smartwatches have affected your real-life friendships?
\item Have you ever had arguments or disagreements with someone because of smartwatches?
\end{itemize}

\subsection{Comparison and Speculative Design}

\subsubsection{Compare with Smartphones}
\begin{itemize}
\item What are the advantages of a smartwatch compared to a smartphone?
\item How does a smartwatch fare in terms of social interactions?
\item When it comes to protecting your privacy, which is more secure, a smartwatch or a smartphone?
\end{itemize}

\subsubsection{Speculative Design}
\begin{itemize}
\item If you could design your own smartwatch, what kind of features would you want it to have?
\end{itemize}

\end{CJK*}
\end{document}